\documentclass[a4paper,11pt]{article}
\pdfoutput=1 

\usepackage{jheppub} 

\usepackage[T1]{fontenc} 

\usepackage[useregional,showdow=true,showzone=false,german]{datetime2}	

\usepackage{lmodern}

\usepackage{xcolor}
\usepackage{arydshln}
\usepackage[font=footnotesize]{caption}
\usepackage{float}
\usepackage{braket}
\usepackage{extarrows}
\usepackage{bbm}
\usepackage{nicefrac}
\usepackage{import}
\usepackage{bbold}
\usepackage{mathrsfs}
\usepackage{pgf,tikz}
\usetikzlibrary{positioning,arrows}
\usetikzlibrary{decorations.pathmorphing}
\usetikzlibrary{decorations.markings}
\usepackage{pgfplots}
\usetikzlibrary{arrows}
\usepackage[fleqn]{mathtools}
\usepackage{graphicx}
\usepackage{subcaption}
\usepackage{dsfont}
\usepackage{enumerate}
\usepackage{enumitem}
\usepackage{booktabs}
\usepackage{siunitx}
\usepackage{array}
\usepackage[absolute]{textpos}
\usepackage{siunitx}
\usepackage{fancyvrb}

\newenvironment{gleichung}{\begin{equation}\begin{aligned}}{\end{aligned}\end{equation}\\ \noindent}
\newenvironment{gleichung*}{\begin{equation*}\begin{aligned}}{\end{aligned}\end{equation*}}

\renewcommand{\sc}{\textsc}
\renewcommand{\bf}{\textbf}

\newcommand{\dd}{\mathrm{d}}

\newcolumntype{?}{!{\vrule width 1.5pt}}
\title{\boldmath Analytic Structure of all Loop Banana Amplitudes}

\author[a]{Kilian B\"onisch}
\author[a]{Fabian Fischbach}
\author[a,b]{Albrecht Klemm}
\author[a,1]{Christoph Nega\note{Corresponding author.}}
\author[a]{Reza Safari}

\affiliation[a]{Bethe Center for Theoretical Physics,\\Universit\"at Bonn, D-53115, Germany}
\affiliation[b]{Hausdorff Center for Mathematics,\\Universit\"at Bonn, D-53115, Germany}

\emailAdd{boenisch@th.physik.uni-bonn.de}
\emailAdd{fischbach@physik.uni-bonn.de}
\emailAdd{aklemm@th.physik.uni-bonn.de}
\emailAdd{cnega@th.physik.uni-bonn.de} 
\emailAdd{rsafari@th.physik.uni-bonn.de}

\abstract{Using the  Gelfand-Kapranov-Zelevinsk\u{\i}  system  for the primitive cohomology 
of an infinite series of complete intersection Calabi-Yau manifolds, whose dimension is the loop
order minus one,   we completely  clarify the  analytic structure of all banana amplitudes with arbitrary 
masses.  In particular, we find that the leading logarithmic structure in the high energy regime, which  corresponds 
to the point of maximal unipotent monodromy,   is determined by a novel $\widehat \Gamma$-class evaluation 
in the ambient spaces of the mirror, while the imaginary part of the amplitude in this regime is determined 
by the $\widehat \Gamma$-class of the mirror Calabi-Yau manifold itself.  We provide simple closed all loop 
formulas  for the former as well as for the Frobenius $\kappa$-constants, which determine the behaviour of the amplitudes, 
when the momentum square equals the sum of the masses squared, in terms of zeta values. We extend our previous work 
from three to four loops by providing for the latter case a complete set of (inhomogenous) Picard-Fuchs differential equations 
for arbitrary masses. This  allows to evaluate the amplitude as well as other master integrals with raised powers 
of the propagators in very short time to very high numerical precision for all values of the physical 
parameters. Using a recent $p$-adic analysis of the periods we determine the value of the maximal cut 
equal mass four-loop amplitude at the attractor points in terms of periods of modular weight two and 
four Hecke eigenforms and the quasiperiods of their meromorphic cousins.    
         
}

\begin{document}
\rightline{
BONN-TH-2020-06
}

\maketitle

\flushbottom


\section{Introduction}
\label{sec:intro}

\noindent Precision calculations of physical observables within a perturbative QFT usually require the evaluation of (many) Feynman integrals up to a certain loop order. This includes, perhaps most prominently, cross sections for high-energy collider experiments. In order to make the most out of an increasing amount of experimental data collected e.g. at the Large Hadron Collider, theoretical predictions need to be sharpened by computing higher loop orders in the perturbative expansion. This poses challenging computational problems that come along with a series of deep questions regarding the mathematical structures underlying multi-loop Feynman integrals.

\begin{figure}[h]
	\centering
	\includegraphics[width=0.6\textwidth]{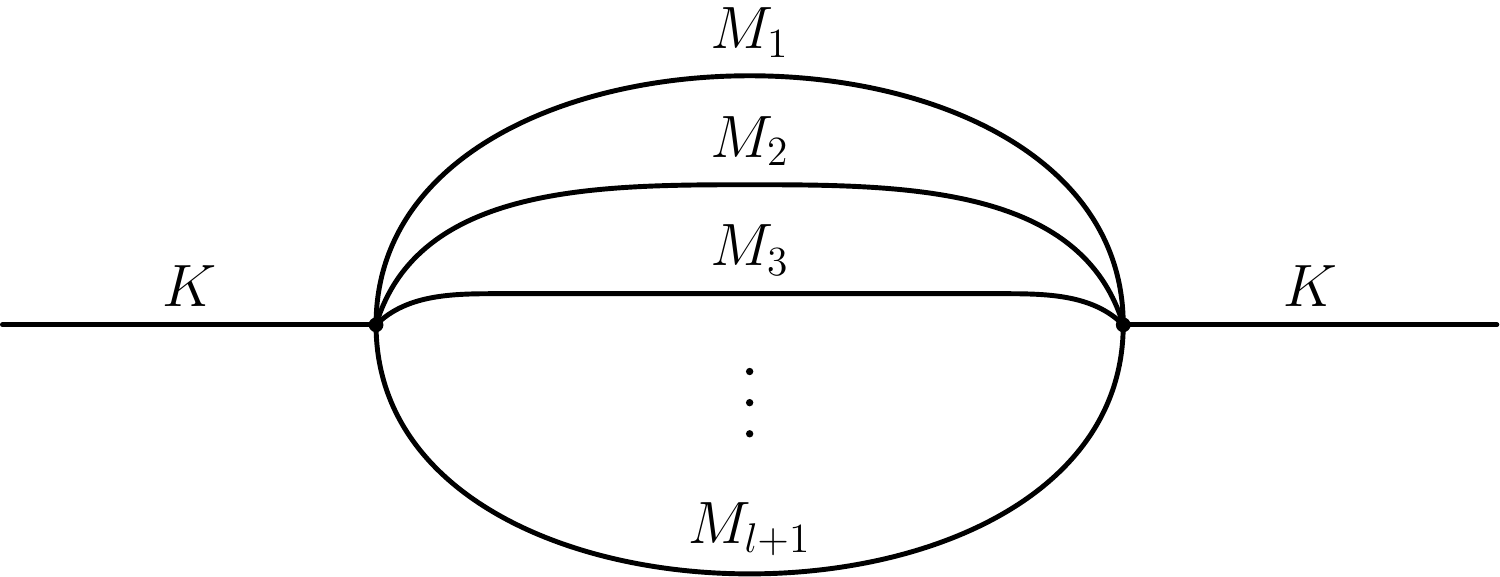}
     \caption{The $l$-loop banana diagram with external momentum $K$ and internal masses $M_i$.}
     \label{figbanana}
\end{figure}

A concrete problem of this kind is defined by the class of Feynman integrals associated with so-called banana graphs that will be considered in the present work (see Figure \ref{figbanana}). Part of their practical relevance comes from the fact that banana type graphs often appear as a subtopology of more complicated (more realistic) Feynman graphs, i.e., they are obtained by contracting a suitable subset of internal lines. To understand this, first note that, after suitable tensor and Dirac algebra manipulations in the numerator of a given Feynman integral (say involving fermion or gauge boson propagators), the problem can generically be reduced to the computation of a set of scalar Feynman integrals, possibly with non-trivial but scalar numerator (in momentum space). Now scalar Feynman integrals often satisfy integration-by-parts identities \cite{Chetyrkin:1981qh,Tkachov:1981wb}\footnote{These identities are reviewed in for example \cite{Grozin:2011mt,Zhang:2016kfo,Smirnov:2012gma}.} (IBP) which allow to further reduce to a smaller (finite) number of integrals, commonly called master integrals of the respective problem. Typically the latter are Feynman integrals associated with subtopologies in the above sense and in this way banana type integrals\footnote{Here the banana type integrals may have propagator powers $\nu_i$ different from unity, in which case one assignes a Feynman graph with $\nu_k -1$ dots on the $k$th propagator.} frequently arise --- for instance as master integrals in two-loop electro-weak computations \cite{Bauberger:1994nk}, in the two-loop Higgs+jet production cross section \cite{Bonciani:2019jyb}, in three-loop corrections to the $\rho$-parameter \cite{Abreu:2019fgk} or at four-loop order in the anomalous magnetic moment of the electron \cite{Laporta:2008sx}.
\tikzset{
photon/.style={decorate, decoration={snake, segment length=2mm, amplitude = 0.5mm}, draw=black},
particle/.style={draw=black, postaction={decorate},
    decoration={markings,mark=at position .5 with {\arrow[draw=black,thick]{>}}}},
antiparticle/.style={draw=black, postaction={decorate},
    decoration={markings,mark=at position .5 with {\arrow[draw=black]{<}}}},
gluon/.style={decorate, draw=black,
    decoration={coil,amplitude=2pt, segment length=5pt}}
 }
\begin{figure}[h]
        \centering
        \begin{tikzpicture}
        
        \draw [gluon] (-1.5,1.5)--(0,1.5);
        \draw [gluon] (-1.5,0)--(0,0);
        
        \draw [particle] (1.5,0)--(0,0);
        \draw [particle] (0,0)--(0,1.5);
        \draw [particle] (0,1.5)--(1.5,1.5);
        \draw [particle] (1.5,1.5)--(1.5,0);
        \draw (0.3,0.75) node {b};
        
        \draw [gluon] (1.5,1.5)--(3,1.5);
        \draw [gluon] (1.5,0)--(3,0);
    
        \draw [particle] (3,1.5)--(3,0);
        \draw [particle] (4.5,0.75)--(3,1.5);
        \draw [particle] (3,0)--(4.5,0.75);
        \draw (3.3,0.75) node {t};
        
        \draw[dashed] (4.5,0.75)--(5.5,0.75);
        \begin{scope}[shift={(5.5,0.75)}]
            \draw (1,0)--(2,0)--(3,1)--(3,-1)--(4,0)--(5,0);
            \draw [photon] (2,0)--(3,-1);
            \draw [photon] (3,1)--(4,0);
        \end{scope}
        
        \end{tikzpicture}
               \caption{A three loop contribution to Higgs production via gluon fusion with a bottom and a top quark running in the loops (left panel). The scalar kite Feynman graph with two massless and three propapagators of equal mass (right panel). }
     \label{fig:feynintro}
\end{figure}
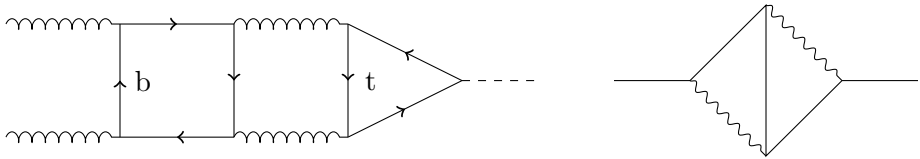
 Another example is the so-called kite integral (see Figure \ref{fig:feynintro}), which gives a contribution to the two-loop electron self-energy. The subtopologies in this example sit in the inhomogeneous term of its first order differential equation \cite{Adams:2016xah}. More generally, the kite family with arbitrary powers of the respective propagators has a set of eight master integrals that satisfy a first-order Fuchsian differential system in the momentum parameter. Here two massive two-loop banana type integrals, once with primitive powers of the propagators and once with a propagator raised to the second power, form together with the two-loop tadpole integral\footnote{Note that the only subtopology of the $l$-loop banana graph is a tadpole of $l$-loop bouqet topology.} a closed differential subsystem \cite{Adams:2016xah}.\footnote{Also for the family of three-loop equal mass banana type integrals one finds 3+1 master integrals, one of which is a tadpole contribution that can be recast to an inhomogeneity in the differential system of the remaining three integrals \cite{Broedel:2019kmn}. See also \cite{Kalmykov:2016lxx,Bitoun:2017nre} for counts of master integrals for banana families.}  Computational control over the subsectors may now be regarded as a key step for solving the full problem. In this spirit explicit analytic results for the three-loop non-equal mass banana Feynman integral are expected to form an important contribution to analytically computing the Higgs production cross section in QCD via gluon fusion with heavy quarks running in the loops, see Figure \ref{fig:feynintro}.\footnote{We thank Claude Duhr for pointing out to us the connection between the three-loop banana integrals and the gluon fusion process depicted in Figure \ref{fig:feynintro}.} 

Higher loop banana integrals also constitute important case studies on the mathematical structure of multi-loop Feynman integrals and the development (or refinement) of appropriate computational tools.
Often Feynman integrals give rise to interesting classes of special functions such as multiple polylogarithms. The latter, however, are not sufficient. In fact, the two-loop banana integral with non-zero masses is known to be the simplest Feynman integral that is not expressible in terms of multiple polylogarithms and it has spurred generalizations to elliptic multiple polylogarithms and iterated integrals of modular forms \cite{Adams:2017ejb,Abreu:2019fgk,Broedel:2019kmn,Ablinger:2017bjx}. Nevertheless, the function theory for higher loop integrals is, especially for different non-zero masses, less understood and explicit results are scarce. The present work, building on \cite{Klemm:2019dbm}, significantly improves this situation for the case of banana type integrals by employing suitable algebro-geometric techniques, well-known in the context of topological string theory on Calabi-Yau manifolds \cite{Hosono:1994ax,MR3965409}. Recall that Feynman integrals, more precisely their Laurent coefficients in the dimensional regularization parameter, are period integrals \cite{Bogner:2007mn} in the sense of Kontsevich and Zagier~\cite{MR1852188}. As a function of their physical parameters (and possibly auxiliary deformation parameters), they satisfy Picard-Fuchs differential equations describing the variation of mixed Hodge structures \cite{MullerStach:2012mp,Vanhove:2014wqa}. Especially, in combination with combinatoric techniques for appropriate toric varieties \cite{MR2810322} these equations become a powerful approach to compute the banana Feynman integrals \cite{Vanhove:2018mto,Klemm:2019dbm}. This eventually allows  for a complete determination of such a Feynman integral in terms of local Frobenius bases of solutions to the Picard-Fuchs differential ideal, which describes a Gelfand-Kapranov-Zelevinsk\u{\i} (GKZ) generalized hypergeometric system \cite{MR1020882,MR1080980,MR1011353}.\footnote{See \cite{NasrollahpPeriodsFeynmanDiagrams2016,Nasrollahpoursamami:2017shc,Feng:2019bdx,Klausen:2019hrg,delaCruz:2019skx} for further applications of GKZ systems and \cite{Schultka:2018nrs,Schultka:2019tfi} for those of toric geometry to Feynman integrals.} 
For the Feynman integral this is an inhomogeneous differential system, the inhomogeneity resulting from the fact that the integration domain has non-trivial boundary so the Feynman integral becomes a relative period.\footnote{Here we refer to the integration domain $\sigma_l$ in the parametric represenation of the Feynman integral, see equations \eqref{bananageneral} and \eqref{sigmal}.} The homogeneous solutions of the system in turn describe period integrals of the same integrand over closed cycles (i.e. without boundary) in the cohomology of a family of Calabi-Yau $(l-1)$-folds, and the maximal cut integral of the Feynman graph turns out to be a special case thereof.\footnote{Also see \cite{Primo:2016ebd,Primo:2017ipr} for a connection between the maximal cut integral and homogeneous differential equations.} In practice this means that the (full) Feynman integral will be a linear combination of the homogeneous solutions (a Frobenius basis for the closed periods) and a special inhomogeneous solution (regarded as a special relative period).\footnote{For simplicity we will sometimes just speak of a Frobenius basis, which then includes the special solution.}

\vspace{3mm}

The paper is organized as follows: In section \ref{sec:banint} we recall the Feynman and Bessel representation of the banana integrals, comment on the underlying Calabi-Yau families and compute the maximal cut integral. In section \ref{sec:anapropequal} we re-derive the equal-mass case differential equation, find Frobenius solutions in the large momentum regime, express the Feynman integral in terms of the latter and discuss monodromies and singularities. The resulting expansion is compared to the $\widehat\Gamma$-conjecture for mirror Calabi-Yau geometries.  Picard-Fuchs operators, multi-variate Frobenius bases and the correct choice of linear combination for the banana integral in the non-equal mass case are provided in section \ref{sec:nonequalmass}, where we also comment on obtaining more general banana type integrals with higher powers of propagators. Section \ref{sec:conclusion} contains conclusions and open problems. Appendix \ref{appbessel} concers the Bessel representation of banana integrals and appendix \ref{sec:appop} deals with an explicit inhomogeneous differential equation for the four-loop generic mass case. The third appendix \ref{sec:pariprogram} explains our \texttt{Pari/GP} script \texttt{BananaAmplitude.gp} which computes the equal mass banana amplitude.

\section{Banana Feynman integrals}
\label{sec:banint}
In this section we introduce the main object we focus on in this paper, namely the $l$-loop banana Feynman integral and make first comments on the underlying geometry. Moreover, we give a representation of the Feynman integral in terms of Bessel functions valid for small momenta. In the large momenta regime we calculate the maximal cut integral.

\subsection{The  $l$-loop banana amplitude and its geometrical realization}
\label{sec:l-loopbanadef}  
The $l$-loop banana Feynman graph is shown in Figure \ref{figbanana}. By means of standard textbook Feynman rules one can write down the Feynman integral associated to each such banana graph. For our purpose it is, however, more convenient to use the Symanzik parametrization of Feynman integrals (see e. g. \cite{Bogner:2010kv} for an introduction). In two dimensions\footnote{By dimensional shift relations \cite{Tarasov:1996br,Lee:2009dh} we can relate the Feynman integral in two dimensions to the leading term in dimensional regularization in four dimensions.} the $l$-loop Feynman integral thus reads
\begin{gleichung}
{\cal F}_{\sigma_l}(t,\xi_{i})=\int_{\sigma _{l}} \dfrac{\mu_{l}}{P_l(t,\xi_i;x)}=\int_{\sigma _{l}} \dfrac{\mu_{l}}{\left(t-\left(\sum_{i=1}^{l+1} \xi_i^2 x_i\right) \left(\sum_{i=1}^{l+1} x_i^{-1}\right) \right)\prod_{i=1}^{l+1} x_i  }~.
\label{bananageneral}
\end{gleichung}
Here the edge variables $x_i$ form a set of homogeneous coordinates for the projective space $\mathbb{P}^l$ and the $l$ real dimensional 
integration domain $\sigma_l$ is defined as 
 \begin{gleichung}
 \sigma_l=\{[ x_1:\ldots: x_{l+1} ] \in \mathbb{P}^l \, | \, x_i \in \mathbb{R}_{\ge 0} \ \text{for all} \ 1 \leq i \leq l+1 \}  \ ,  
\label{sigmal}
\end{gleichung}  
while the holomorphic $l$-measure $\mu_l$ is defined by
\begin{gleichung}
\mu_{l} = \sum_{k=1}^{l+1} (-1)^ {k+1} x_k \ \dd x_1 \wedge  \ldots \wedge {\widehat {\dd x_k} } \wedge  \ldots \wedge \dd x_{l+1}~ .
\label{measure}
\end{gleichung}
As usual, the hat indicates the omission of one differential.  In writing \eqref{bananageneral} we have also introduced dimensionless kinematical parameters,
\begin{equation}
t =\frac{K^2}{\mu^2}\, , \qquad \text{and} \qquad \xi_i = \frac{M_i}{\mu} \qquad (i=1, \ldots l+1) \, ,
\end{equation}
where $K$ is the external momentum, the $M_i$ are the $l+1$ propagator masses and $\mu$ is an arbitrary infrared scale.

A key observation  is that \eqref{bananageneral} can be understood as a (relative-) period 
integral for a smooth family of   Calabi-Yau hypersurfaces  $M_{l-1}^{l^2}$  with generically ${\rm dim} {\, H^1(M_{l-1}^{l^2},TM_{l-1})}=h^{l-2,1}=l^2$ 
complex structure deformations\footnote{See \cite{Klemm:2019dbm} for a more detailed description of  
the reflexive pair of lattice polyhedra  $(\Delta_l,\hat \Delta_l)$ and the associated almost Fano--  ($\mathbb{P}_{\Delta_l},\mathbb{P}_{\hat \Delta_l})$  and Calabi-Yau (mirror) geometries $(M_{l-1},{\widehat {M_{l-1}}})$.}       
\begin{equation}  
M_{l-1}^{l^2}=\{P_{\Delta_l}({\underline y})=0 \, | \, {\underline y} \in  \mathbb{P}_{\hat \Delta_l} \} \ , 
\label{CTMl}  
\end{equation}  
defined as vanishing locus of the Laurent polynomial $P_{\Delta_l}=P_l(t,\xi_i;x)/\prod_{i=1}^{l+1} x_i$ in the 
coordinate ring of the toric ambient space $\mathbb{P}_{\hat \Delta_l}$, where $\Delta_l$ is the $l$-dimensional reflexive Newton polytope of the 
polynomial $P_{\Delta_l}$ and $\hat \Delta_l$ is its dual.  By Batyrev's mirror construction~\cite{MR1269718} the mirror $W_{l-1}$ 
is given as in (\ref{CTMl}) but with the r\^ole  of $\Delta_l$ and $\hat \Delta_l$ exchanged.  

We note that a  single residuum integral  $\Omega_{l-1}={\rm Res}_{P_l(t,\xi_i;x)=0} \left(\mu_l/P_l(t,\xi_i;x)\right)$ yields an 
expression for the holomorphic 
$(l-1,0)$-form on the Calabi-Yau manifolds $M_{l-1}^{l^2}$. This was used in \cite{Klemm:2019dbm} to derive  
from the  Gelfand-Kapranov-Zelevinsk\u{\i} (GKZ) differential system up to three loops the differential D-module
describing those geometrical integrals  over  $\Omega_{l-1}$ that yield the physical Feynman 
amplitude  (\ref{bananageneral}) in all regions of their  physical parameter space  in $t$ and $\xi_i$ for $i=1,\ldots, l+1$.

In the present work we extend this program~\cite{Klemm:2019dbm}  to banana Feynman diagrams of all loop 
orders $l$. As in~\cite{Klemm:2019dbm} a key technical step is  to reduce the solutions of the $l^2$ 
parameter GKZ system to the subset of solutions  that describe the physical periods in the 
$l+1$ physical parameters, which we achieved  starting from the GKZ system of $M_{l-1}^{l^2}$. 
Even though the full differential D-module   ${\cal D}^l$ is lengthy  to write down and will be 
made explicit  only up to $l=4$, we can provide a complete   analytic description for the amplitude for  
the $l$-loop banana graph ${\cal F}_{\sigma_l}(t,\xi_{i})$  in all regions of the moduli space.  
The latter is based on the identification  of certain  universal operators in ${\cal D}^l$ and 
systematic analytic continuation formulas valid  for all $l$, which involves a systematic 
occurrence of products of zeta values with highest transcendentality $l$. It is natural to 
expect that the  latter are related to the $\widehat \Gamma$-class of the  mirror Fano threefold 
$\mathbb{P}_{\hat \Delta_l}$. Due to the very high co-dimension of the K\"ahler 
subslice dual to the $(l+1)$-dimensional physical slice of parameters of $M_{l-1}^{l^2}$ this is an 
increasingly complicated task.  However,  Matt Kerr pointed out to us  that for the Fano variety that is
associated to the equal mass three-loop banana diagram there is a realization of its Hodge structure 
that is an alternative to the redundantly parametrized one of $\mathbb{P}_{\hat \Delta_l}$ and is  
simply a  degree $(1,1,1, 1)$ hypersurface  in $(\mathbb{P}^1)^4$. This key observation was made by 
identifying  the holomorphic solution associated to the differential system of this amplitude 
with the one that appears in an example\footnote{This example is given in the link \url{http://coates.ma.ic.ac.uk/fanosearch/?page\_id=277\#4-1}.}  in the  list \cite{Fanosearch} and has the same description. This suggests that the  relevant physical 
subslices in the series of the Calabi-Yau  manifolds $M_{l-1}^{l^2}$ (\ref{CTMl})   are complete intersections of   two degree 
$(1,\ldots,1)$ constraints in $(\mathbb{P}^1)^{l+1}$. The GKZ systems of complete intersections 
have been studied in~\cite{Hosono:1994ax} under the aspect of mirror symmetry. So a  good model for the 
Calabi-Yau $(l-1)$-fold $W_{l-1}$ is the complete intersection of two degree $(1,\ldots,1)$ constraints in $(\mathbb{P}^1)^{l+1}$ that reads in the notation 
of~\cite{Hosono:1994ax}
\begin{equation} 
W^l_{l-1}=
\left(\begin{array}{c} \mathbb{P}^1_1\\ \vdots \\      \mathbb{P}^1_{l+1}\end{array}\right|\!\! \left|  \left. \begin{array}{cc} 1&1 \\ \vdots &\vdots \\ 1& 1\end{array}\right\} \ l+1\right)\subset    
\left(\begin{array}{c} \mathbb{P}^1_1\\ \vdots \\      \mathbb{P}^1_{l+1}\end{array}\right|\!\! \left|  \left. \begin{array}{c} 1\\ \vdots  \\1\end{array}\right\} \ l+1\right)=F_l\ ,
\label{CICY} 
\end{equation} 
which is here suitably embedded\footnote{The lower index on the  manifolds (apart from $\mathbb P_k^1$ of course) indicates their complex dimension in terms of the number of loops $l$ of the 
Feyman diagram.}    in the Fano $l$-fold $F_l$. According  to~\cite{Hosono:1994ax}  the 
mirror manifold $M_{l-1}^{l+1}$ is given by a resolved quotient of (\ref{CICY}) $M_{l-1}={\widehat {W_{l-1}/G} }$  and the period  geometry of 
$M_{l-1}^{l+1}$ is defined by the invariant periods of $W_{l-1}/ G$ depending on the $G$ invariant  $l$-dimensional  
deformation space.  This construction is a special case of the construction of Batyrev and  Borisov~\cite{Batyrev:1994pg}.

This suggests that the physical mass and momentum parameters  should be identified  in the high 
energy regime with the complexified  K\"ahler parameters  
\begin{equation}
\mathfrak t^k=\frac{1}{2\pi i} \int_{\mathbb{P}_k^1} (i\omega -  b)
\end{equation} 
 controlling  the area\footnote{Here $\omega$ is K\"ahler form and  the complexification is by the expectation 
 value of the Neveu-Schwarz $(1,1)$-form field $b$.}  $A_k=\frac{1}{2 \pi} \int_{\mathbb{P}^1_k} \omega$ 
 of the $k$th  $\mathbb{P}^1$ in (\ref{CICY}) as 
\begin{equation} 
\mathfrak t^k\simeq \frac{1}{2 \pi i} \log\left(\frac{M_k^2}{K^2} \right)=\frac{1}{2 \pi i}\log(z_k)  \qquad \text{for }k=1,\ldots , l+1 \ . 
\end{equation}         
If this is true we expect that the large energy behaviour of the  Feynman amplitude   
is  exactly determined by the quantum cohomology  of $W_{l-1}\subset F_l$ in the large 
volume limit of the geometry. In particular, if  this beautiful picture holds we can 
infer the entire  leading logarithmic structure  of the Feynman graph from the central 
charge of the corresponding object in the derived category of coherent sheafs 
which  can be described  by the  $\widehat \Gamma$-class conjecture in terms of the 
topological data of  $W_{l-1}$  as well as of  $F_l$, which can be easily controlled 
for all $l$.  Here $z_k$ are the canonical complex structure variables of  $M_{l-1}$, chosen 
so~\cite{Hosono:1994ax} that the point of maximal unipotent monodromy of the 
Picard-Fuchs-- (or Gauss-Manin) system of $M_{l-1}$   is at $z_k=0$. We establish 
the equivalence of the two geometric descriptions, by first  deriving the Picard-Fuchs equations of the
Feynman graph geometry  \eqref{bananageneral} and \eqref{CTMl} as Calabi-Yau hypersurface  in a 
toric variety by reduction of its GKZ system  to the physical  parameters  and finding their solutions. 
These data can be compared to the GKZ system for the complete intersection   (\ref{CICY}) 
and its solutions given in~\cite{Hosono:1994ax} after a change of variables. Note, however, that 
the GKZ systems  given  in~\cite{Hosono:1994ax} in generality do not yield immediately the 
complete Picard-Fuchs differential ideal for closed Calabi-Yau periods which entirely the maximal cut case. 
We solved this problem for the homogenous system for the Calabi-Yau periods and the extension 
to the inhomogenous  system for the three-loop graph in~\cite{Klemm:2019dbm}  and for the four-loop 
graph in this work. In the general loop case we can check that the holomorphic solutions (\ref{maxcutmum}) and 
(\ref{defcn}) that can be in both geometries derived  from a simple residuum integral near the MUM point 
agree with a suitable identification of the variables.

\subsection{Bessel function representation of $l$-loop banana integrals}
Besides the parametric representation \eqref{bananageneral}, we also recall a representation of the Feynman amplitude in terms of an integral over Bessel functions, which in its regime of validity,
\begin{gleichung}
	t	<	\left(\sum_{i=1}^{l+1}\xi_i\right)^2~,
\end{gleichung}
 is well suited for numerical evaluation. Relegating a short derivation to appendix~\ref{appbessel}, the Feynman integral \eqref{bananageneral} can be rewritten as
\begin{gleichung}
	{\cal F}_{\sigma_l}	=	2^l\int_0^\infty z \, I_0(\sqrt t z) \prod_{i=1}^{l+1}K_0(\xi_i z)~ \dd z~.
\label{besselform}
\end{gleichung}	
In particular, in the equal mass case the expression \eqref{besselform} contains the $(l+1)$th symmetric power of the Bessel function $K_0$.

As a side remark, in the on-shell case (defined via $t=\xi_i=1$ for all $i$) the integral \eqref{besselform} becomes a special instance of a Bessel moment. Bessel moments differ in their powers of $z$, $I_0(z)$ and $K_0(z)$ in the integrand. The massive vacuum banana integrals also yield Bessel moments \cite{bailey2008elliptic,Broadhurst2013}. Such Bessel moments have also caught the interest of number theorists, one reason being that they, in some cases, evaluate to critical values of L-series of certain modular forms (as briefly reviewed in subsection \ref{subsec:specialpoints}). They satify many interesting relations \cite{Broadhurst:2018tey,Zhou:2017jnm,Zhou:2018tva,Zhou:2019rgc} and are closely related to L-functions built from symmetric power moments of Kloosterman sums \cite{Broadhurst:2016myo,fresn2018hodge,fresn2020quadratic}.

\subsection{The maximal cut integral for large momentum}
One goal of the paper is to analyze the $l$-loop banana graph in the regime $t>\left(\sum_{i=1}^{l+1}\xi_i\right)^2$ where the expression \eqref{besselform} becomes invalid. It turns out that there is an elegant expression for the so-called maximal cut integral associated with the banana graph, which still contains substantial information about the full Feynman integral ${\cal F}_{\sigma_l}$.

The maximal cut integral is obtained by replacing all\footnote{One can also consider non-maximal cuts where only some propagators are replaced by delta functions. For our purpose these cut integrals are not relevant.} propagators by delta functions. As derived in \cite{Vanhove:2018mto} there is again a parametric representation of the maximal cut integral in terms of the Symanzik polynomials. To get the maximal cut integral one simply changes the integration range from the simplex $\sigma_l$ to the $l$-torus $T^l$. So for the banana integrals we obtain
\begin{gleichung}
	\mathcal F_{T^l}(t,\xi_i)		=	\int_{T^l} \dfrac{\mu_{l}}{\left(t-\left(\sum_{i=1}^{l+1} \xi_i^2 x_i\right) \left(\sum_{i=1}^{l+1} x_i^{-1}\right) \right)\prod_{i=1}^{l+1} x_i  }~.
\label{maxcutdef}
\end{gleichung}

Now, for large momenta $t$, the maximal cut integral $\mathcal F_{T^l}$ can be obtained explicitly by a simple residue calculation. Introduce the variable $s=1/t$, then for small $s$ subsequent geometric series and multinomial expansion yields
\begin{gleichung}
	\mathcal F_{T^l}(t,\xi_i)	&=	\int_{T^l} \frac{s}{1-s\left(\sum_{i=1}^{l+1} \xi_i^2 x_i\right) \left(\sum_{i=1}^{l+1} x_i^{-1}\right)} ~ \frac{\mu_l}{\prod_{i=1}^{l+1}x_i}	\\
						&=	\int_{T^l} \sum_{n=0}^\infty s^{n+1} \sum_{|k|=n}\binom{n}{k_1,\hdots,k_{l+1}}\prod_{i=1}^{l+1}(\xi_i^2x_i)^{k_i}\\ 
						&\qquad \cdot\sum_{|\tilde k|=n}\binom{n}{\tilde k_1,\hdots,\tilde k_{l+1}}\prod_{i=1}^{l+1} x_i^{-\tilde k_i} ~ \frac{\mu_l}{\prod_{i=1}^{l+1}x_i}	\\
						&=	(2\pi i)^{l}\sum_{n=0}^\infty s^{n+1} \sum_{|k|=n} \binom{n}{k_1,\hdots,k_{l+1}}^2 ~\prod_{i=1}^{l+1}\xi_i^{2k_i}~.
\end{gleichung}
Here we used the short hand notation $|k|=\sum_{i=1}^{l+1}k_i$ and  evaluated a multidimensional residue in the last step. So up to normalization the maximal cut integral is for large momentum given  by
\begin{gleichung}
	\varpi_0(s,\xi_i)	=		\sum_{n=0}^\infty s^{n+1} \sum_{|k|=n} \binom{n}{k_1,\hdots,k_{l+1}}^2 ~\prod_{i=1}^{l+1}\xi_i^{2k_i}~.
\label{maxcutmum}	
\end{gleichung}
When expressed in terms of the variable $t=1/s$, one recovers \cite[eq. (123)]{Vanhove:2018mto}.

\section{The $l$-loop equal mass banana Feynman integral}
\label{sec:anapropequal}
In this section we focus on the equal mass case, i.e. $\xi_i=1$ for $i=1,\hdots l+1$. We first derive an inhomogeneous differential equation for the equal mass Feynman integral. This equation is already known in the literature~\cite{verrill2004sums,Vanhove:2014wqa}, however, the derivation presented here is somewhat different. It is based on the observation that the maximal cut integral in the large momentum regime is given by the double Borel sum of a certain function for which one can easily construct an operator that annihilates it. This relation makes the computation of the desired differential equation conceptually clear and easy. For each $l$ the differential equation thus obtained is related to the one in ~\cite{Vanhove:2014wqa} by simply transforming it to the small momentum regime. We subquently explain the analytic properties of its solutions and compare to the actual Feynman integral. Coefficients relating local solutions in the large momentum regime to the Feynman integral are given. A conjectural relation  for these coefficients to the so-called $\widehat\Gamma$-class is proposed. Moreover, these coefficients are linked to the Frobenius $\kappa$-constants as we explain before ending the section with some remarks about special points of the Feynman amplitude.

\subsection{Inhomogeneous differential equation for the $l$-loop equal mass banana Feynman integral} \label{ssc:pfeequalmass}
In this subsection we give an elegant description for the $l$-loop banana Feynman integral which easily leads to its inhomogeneous differential equation.  

First consider the maximal cut integral $\mathcal F_{T^l}$ for large momenta $t$, i.e. near $s=0$, as given in equation \eqref{maxcutmum}. In the equal mass case, i.e. $\xi_i=1$, the expression $\varpi_0/s$ can be seen as the double Borel sum\footnote{Or in other words, the $(l+1){\text{th}}$ symmetric power of $I_0$ is the Borel transform of the Borel transform of $\varpi/s$.} of the $(l+1){\text{th}}$ symmetric power of the series
\begin{equation}
	\sum_{k=0}^\infty \frac{1}{(k!)^2}x^k	=	I_0(2\sqrt{x})~.
\label{boreltrans}
\end{equation}
Note that the double Borel sum resides, simply speaking, in the additional factor of $(n!)^2$ in the coefficients of \eqref{maxcutmum}, relative to those of the symmetric power of the Bessel function.

Hence, the differential equation annihilating the maximal cut integral \eqref{maxcutmum} can be derived by three steps: First calculate the differential equation for the $(l+1){\text{th}}$ symmetric power of \eqref{boreltrans}. Second, by a simple analysis of the (double) Borel sum we can infer the differential operator of the function $\varpi_0/s$ from the operator of the symmetric power. 
Third, the additional factor of $s$ is commuted into the differential operator to obtain the Picard-Fuchs equation for $\varpi_0$.

\textit{Step 1.} The function $I_0(2\sqrt x)$ is annihilated by the operator
\begin{equation}
	\mathcal D	=	\theta^2-x
\end{equation}
with the logarithmic derivative $\theta = x\hspace{0.4ex}\partial_x$. For the $(l+1)$th symmetric power of this function we use a result from \cite{Borwein_2008,MR1809982}:
\begin{itemize}
\item[ ]\bf{Lemma:} Let $\mathcal D=\theta^2+a(x)\theta+b(x)$ be a linear differential operator whose coefficients $a(x)$ and $b(x)$ are rational functions. Let $\mathcal L_0=1$, $\mathcal L_1=\theta$ and for $k=1,2,\hdots, n$ define the operator $\mathcal L_{k+1}$ by
\begin{equation}
	\mathcal L_{k+1}	=	(\theta+ka(x))\mathcal L_k	+	kb(x)(n-k+1)\mathcal L_{k-1}~.
\end{equation}
Then the symmetric power $y^n$ of any solution to $\mathcal D y=0$ is annihilated by 
$\mathcal L_{n+1}$.
\end{itemize}
In the case at hand we have $a(x)=0$ and $b(x)=-x$ while $n=l+1$.

 \textit{Step 2.} For the Borel summation we notice the following properties: Given a power series $\Psi(x) = \sum_n a_n x^n$, its Borel transform is defined by $\mathcal B\Psi(z)=\sum_n a_n \frac{z^n}{n!}$. The original series $\Psi(x)$ is obtained from the Borel transform $\mathcal B\Psi(z)$ by the back-transformation
\begin{equation}
	\Psi(x)	=	\int_0^\infty \mathrm e^{-z}\hspace{0.5ex}\mathcal B\Psi(zx)~\mathrm dz~,
\label{laptrans}
\end{equation}
which is similar to a Laplace transformation, the right hand side now being referred to as the Borel sum of $\Psi(x)$. 
Given a differential operator annihilating the Borel transform $\mathcal B\Psi(z)$ we can infer the corresponding operator annihilating the original function $\Psi(x)$, simply by analyzing the relation \eqref{laptrans}. The rules
\begin{gleichung}
	\theta_z^n \mathcal B\Psi(z)	&\longrightarrow	\theta_x^n \Psi(x)	\\
	z^n\mathcal B\Psi(z)			&\longrightarrow	(x(1+\theta_x))^n \Psi(x) =  \theta_x\left(x^n\prod_{k=1}^{n-1}(\theta_x+k)\right)\Psi(x)
\label{borelrules}
\end{gleichung}
are useful in this respect, where $\theta_{x,z}$ are the logarithmic derivatives in $x$ and $z$, respectively. After each back-transformation, i.e., application of the rules \eqref{borelrules}, we can factor out a logarithmic derivative $\theta_x$ since it turns out that the degree of the differential operator for the Borel transform of $\varpi_0/s$ is increased by one compared to the original function $\varpi_0/s$.

\textit{Step 3.} Finally, we remark that given a function $f(x)$ and an operator $\mathcal D$ with $\mathcal Df=0$ the function $\phi(x)=xf(x)$ is annihilated by the operator $\tilde{\mathcal D}$, which is obtained from $\mathcal D$ by replacing $\theta \rightarrow \theta-1$.

Putting all together we obtain the homogeneous degree $l$ operator $\mathcal L_l$  annihilating the equal mass maximal cut integral $\mathcal F_{T^l}$. It turns out that this operator is of Fuchsian type for any $l$. Using a computer algebra program such as \texttt{mathematica} it is not hard to write a small program\footnote{On the webpage \url{http://www.th.physik.uni-bonn.de/Groups/Klemm/data.php} we upload a small \texttt{mathematica} file including a program to generate these operators. They are normalized that they start with one.} to generate the differential operators. The first few are listed in Table~\ref{listopmum}.

For the full equal mass banana Feynman integral $\mathcal F_{\sigma_l}$ we have to extend these differential equations to inhomogeneous ones. By numerical evaluation of the integral $\mathcal L_l \mathcal F_{\sigma_l}$ one finds for the inhomogeneity 
\begin{equation}
	\mathcal L_l \mathcal F_{\sigma_l}(s,1)	=	S_l	:=	-(l+1)!\hspace{0.5ex}s~.
\label{eq:pfeeqmass}
\end{equation}
\begin{table}[]
\centering
{\small{
\begin{tabular}{@{}cp{12cm}@{}}
\toprule
$\# $Loops $l$& Differential operator $\mathcal L_l$ \\ \midrule
1 & $1-2 s+(-1+4 s) \theta$\\[1ex]
2 & $1-3 s+(-2+10 s) \theta +(-1+s) (-1+9 s) \theta ^2$\\[1ex]
3 & $1-4 s+(-3+18 s) \theta +(3-30 s) \theta ^2-(-1+4 s) (-1+16 s) \theta ^3$\\[1ex]
4 & $1-5 s+(-4+28 s) \theta +\left(6-63 s+26 s^2-225 s^3\right) \theta ^2+\left(-4+70 s-450 s^3\right) \theta ^3 \newline-(-1+s) (-1+9 s) (-1+25 s) \theta ^4$\\[1ex]
5 & $1-6 s+(-5+40 s) \theta +\left(10-112 s+1152 s^3\right) \theta ^2\newline+\left(-10+168 s-236 s^2+4608 s^3\right) \theta ^3+\left(5-140 s+5760 s^3\right) \theta
   ^4\newline+(-1+4 s) (-1+16 s) (-1+36 s) \theta ^5$\\[1ex]
 \bottomrule
\end{tabular}
}}
\caption{Homogeneous differential operators for maximal cut integrals}
\label{listopmum}
\end{table}

\subsection{Analytic properties of the $l$-loop equal mass Banana  graph Feynman integral}
In this subsection we study the analytic properties of the Frobenius basis corresponding to the (in-)homogeneous differential equaion \eqref{eq:pfeeqmass} derived in the previous subsection. These properties partially descend to the actual Feynman integral, which is given by an appropriate linear combination. The coefficients of this linear combination will be computed in the next subsection.

We reserve the indices $k=0,\hdots,l-1$ to the homogeneous  solutions of $\mathcal{L}_l \varpi_k=0$, while the index $k=l$ refers to the special solution of the inhomogeneous equation $\mathcal{L}_l \varpi_l=S_l$. In this notation the Feynman amplitude $\mathcal{F}_{\sigma_l}$ is a linear combination 
of the $\varpi_k$ with a non-zero contribution of the special solution $\varpi_l$. On the other hand, the maximal cut $\mathcal{F}_{T^l}$ of the Feynman amplitude only involves the homogeneous solutions $\varpi_k$ with $k=0,\ldots,l-1$. 

First, we discuss the singular points of the differential equation. At $s=1/t=0$ we have a point of maximal unipotent monodromy, in short a MUM point. This means that the local exponents (i.e. the roots of the indicial equation) of $\mathcal L_l \varpi_k=0$ are all degenerate. In the case at hand they are all equal to one, which can be derived from the fact that $\mathcal L_l= (1-\theta)^l + \mathcal O(s)$. Moreover, in the $s$ coordinate the singular loci are the roots of the discriminant  $\Delta(\mathcal{L}_l)$, given by
\begin{gleichung}
	\Delta(\mathcal L_l)	=  s \prod_{j=0}^{\lfloor \frac{l+1}{2}\rfloor}\left(1-s(l+1-2j)^2\right)\ . 
\label{discriminantequal}
\end{gleichung}
So in general, we have a moduli space 
\begin{equation} 
\mathbb{P}^1\setminus \left(  \bigcup_{j=0}^{\lceil \frac{l-1}{2}\rceil}  \left\{ \frac{1}{(l+1-2j)^2}\right\} \cup \{0,\infty \} \right) \ ~.
\end{equation} 

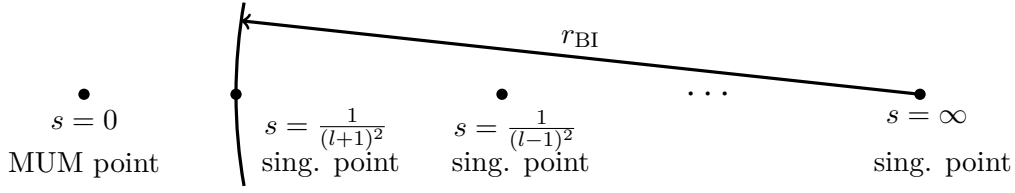
\begin{figure}[h]
    \centering
    \begin{tikzpicture}
    \draw (0,0) node[circle,fill,inner sep=1.5pt,label=below:{$s=0$}] {};
    \draw (0,-.5) node[label=below:{MUM point}] {};
    \draw (2,0) node[circle,fill,inner sep=1.5pt,label=below right:{\hspace{5pt}$s=\frac{1}{(l+1)^2}$}] {};
    \draw (3.25,-.5) node[label=below:{sing. point}] {};
    \draw (5.5,0) node[circle,fill,inner sep=1.5pt,label=below:{\hspace{10pt}$s=\frac{1}{(l-1)^2}$}] {};
    \draw (5.75,-.5) node[label=below:{sing. point}] {};
    \draw (8.25,0) node {{\Large$\cdots$}};
    \draw (11,0) node[circle,fill,inner sep=1.5pt,label=below:{\hspace{5pt}$s=\infty$}] {};
    \draw (11.3,-.5) node[label=below:{sing. point}] {};
    \draw[very thick] (2,0) arc (180:170:7);
    \draw[very thick] (2,0) arc (180:190:7);
    \draw [very thick][->] (11,0)--({9+7*cos(172)},{7*sin(172)});
    \draw ({10+3.5*cos(172)},{3.5*sin(172)}) node [above] {$r_{\mathrm{BI}}$};
    \end{tikzpicture}
    \caption{Singularities of the Fuchsian operator $\mathcal{L}_l$. The radius of convergence of the Bessel integral representation \eqref{besselform} of $\mathcal{F}_{\sigma_l}$ is denoted by $r_{\mathrm{BI}}$.}
    \label{fig:ModuliSpace}
\end{figure}

The actual Feynman integral $\mathcal F_{\sigma_l}$ is not singular at all of these points.
From the Bessel function  representation \eqref{besselform} of $\mathcal{F}_{\sigma_l}$, valid in a neighbourhood of the point $s=\infty$ (i.e. $t=0$) we know that it converges for $|s| > \frac{1}{(l+1)^2}$. In particular, this implies that the amplitude can only be singular at $s=0$ and $s=\frac{1}{(l+1)^2}$. This is also expected from the optical theorem since the latter point is a threshold for the $l+1$ particles in the loops (all of which have unit mass). At the other singular points of $\mathcal L_l$ the Feynman integral stays regular. In Figure \ref{fig:ModuliSpace} this behavior is shown.

\subsection{Frobenius basis at the MUM point} 
\label{ssc:frobeniusEM}
Around the MUM point the Frobenius basis takes a particularly nice form. The holomorphic solution $\varpi_0$ is given by \eqref{maxcutmum} at $\xi_i=1$ for $i=1,\hdots,l+1$. The other solutions are given by\footnote{Note that the dependence on the loop order $l$ is kept implicit in our notation for $\varpi_k$  and $\Sigma_k$.} 
\begin{gleichung}
	\varpi_k	=	\sum_{j=0}^k \binom{k}{j}\log(s)^j \, \Sigma_{k-j}	\qquad\text{for }k=1,\hdots,l-1~,
\end{gleichung}
where $\Sigma_0=\varpi_0 = s + \mathcal O(s^2)$ and the power series $\Sigma_k$ are determined by the operator $\mathcal L_l$ and the condition that they start as $\Sigma_k = \mathcal O(s^2)$ for $k\geq1$. For example, the four-loop operator $\mathcal L_4$ has
\begin{gleichung}
	\varpi_0	&=	s + 5s^2 + +45s^3 + 545s^4 + \num{7885}s^5 + \cdots	\\
	\Sigma_1	&=	8s^2 + 100s^3 + \frac{\num{4148}}3s^4 + \frac{\num{64198}}3s^5 + \cdots	\\
	\Sigma_2	&=	2s^2 + \frac{197}2s^3+\frac{\num{33637}}{18}s^4 + \frac{\num{2402477}}{72}s^5 + \cdots	\\
	\Sigma_3	&=	-12s^2 - \frac{267}2s^3 - \frac{\num{19295}}{18}s^4 - \frac{\num{933155}}{\num{144}}s^5 + \cdots~.
\end{gleichung}
The special solution $\varpi_l$ has one more logarithm and takes the form
\begin{gleichung}
	\varpi_l	=	\sum_{j=0}^l \binom{l}{j}\log(s)^j \, \Sigma_{l-j}~,
\end{gleichung}
which, after multiplication with the constant $(-1)^{l+1}(l+1)$, satisfies \eqref{eq:pfeeqmass}. Again for the four-loop example we find
\begin{equation}
	\Sigma_{4}	=	\num{1830}s^3 + \frac{\num{112720}}3s^4 + \frac{\num{47200115}}{72}s^5 + \cdots ~
\end{equation}
in the special solution $\varpi_4$.

The power series $\Sigma_k$ can also be obtained from a generating function approach starting from the holomorphic solution $\varpi_0$. To this end rewrite \eqref{maxcutmum} in the form
\begin{gleichung}
	\varpi_0	=	\sum_{ k_1, \dots , k_{l+1} \geq 0 } \binom{|k|}{k_1,\hdots,k_{l+1}}^2 ~s^{|k|+1}~,
\label{varpi0anders}
\end{gleichung}
where the summation is over all non-negative integers $k_1,\hdots,k_{l+1}$. 
Introduce formal parameters $\epsilon_i$ by replacing $k_i\rightarrow k_i + \epsilon_i$ in \eqref{varpi0anders}. Taking derivatives with respect to these parameters and subsequently putting the parameters to zero yields other Frobenius solutions\footnote{The functions produced here may be linear combinations of the previously considered solutions to the homogeneous differential equation. In other words, by forming appropriate linear combinations of the expressions in \eqref{eq:equaldiffeps} one in turn obtains the $\Sigma_k$ as defined before. }
\begin{gleichung}\label{eq:equaldiffeps}
		\sum_{ \lbrace i_1,\hdots, i_k \rbrace \in T_k^{(l+1)}}\left.\frac{\partial^k}{\partial\epsilon_{i_1}\cdots\partial\epsilon_{i_k}}\varpi_0(\epsilon_i)\right|_{\text{all} \ \epsilon_i=0}	\quad\text{for } k=1, \hdots, l~,
\end{gleichung}
where $T_k^{(l+1)}$ denotes the set of all subsets of length $k$ of the set $ T^{(l+1)}=\{1,\hdots,l+1\}$.


\subsection{Banana Feynman integral in terms of the MUM-Frobenius basis}\label{ssc:linearcombi} 

Recall that in a local Frobenius basis (for us the region around the MUM point is most interesting) the $l$-loop banana Feynman amplitude is given by a linear combination
\begin{equation}\label{eq:deflambda}
	\mathcal F_{\sigma_l}	=	\sum_{k=0}^{l} \lambda_k^{(l)} \varpi_k~.
\end{equation}
In the following we explain how to obtain the coefficients $\lambda_k^{(l)}$.


The equal mass banana Feynman integrals 
$\mathcal F_{\sigma_l}$ can be evaluated numerically for fixed value of the variable $s$. For $l=2,3,4$ this can directly be done with the form given in \eqref{bananageneral}, say using a numerical integration routine of \texttt{mathematica} or \texttt{pari}. Unfortunately, the multidimensional numerical integration gets too cumbersome for higher loop integrals due to the increase of the numerical error. On the other side, numerical integration is less problematic for the integral over Bessel functions \eqref{besselform}, which can almost be computed for any loop order with any desired precision. However, around the MUM point $s=0$ the Bessel expression is not valid and analytic continuation is needed. As seen from Figure~\ref{fig:ModuliSpace} we therefore have to analytically continue the solutions from $s=\tfrac{1}{(l+1)^2}$ to $s=0$. For this, the Bessel representation of $\mathcal F_{\sigma_l}$ is used to first fix the linear combination with respect to the local Frobenius basis\footnote{These are obtained by shifting the variable $s$ to $\eta=s-\tfrac{1}{(l+1)^2}$ and solving the differential equation around $\eta=0$. It turns out that one obtains square root or logarithmic brunch cuts depending on whether $l$ is odd or even, respectively.} around $s=\tfrac{1}{(l+1)^2}$. Then, by subsequent numerical analytic continuation, the local Frobenius basis around $s=\tfrac{1}{(l+1)^2}$ is related to the local Frobenius basis at the MUM point $s=0$. This numerically yields the desired coefficients $\lambda_k^{(l)}$ around $s=0$.

In order to guess the exact analytic expression of the coefficient $\lambda_k^{(l)}$ we take as an ansatz\footnote{This ansatz is inspired by the $\widehat\Gamma$-conjecture, which will be addressed in the next subsection.} all possible products of zeta values and $\pi$ that lead to a homogeneous transcendental weight of $l-k$ and linearly combine these products with rational coefficients to be determined. The latter are fitted by comparing the ansatz to the numerical values for $\lambda_k^{(l)}$ obtained in the previous paragraph.\footnote{If we write down the ansatz with even powers of $\pi$ instead of even zeta values the coefficients in the ansatz turn out to be integers.} 
 We checked these fits with 300 digits precision up to the loop order $l=17$ and with lower precision until $l=20$. For example, in the four-loop case we thus find (see also Table \ref{tablambdaequal})
\begin{gleichung}
	\lambda_0	^{(4)}	&=	\num{-450}\zeta(4) - i\pi\cdot	80\zeta(3)	\qquad	&\lambda_1	^{(4)}	&=	\num{80}\zeta(3) - i\pi\cdot	120\zeta(2)	\\
	\lambda_2	^{(4)}	&=	\num{180}\zeta(2)	\qquad	&\lambda_3^{(4)}	&=	i\pi\cdot 20	\hspace{3cm} 	\lambda_4	^{(4)}	=	-5~.
\end{gleichung}
In all examples considered we observe that the imaginary part factors into $\pi$ and a sum of homogeneous transcendental weight $l-k-1$.

Moreover, all empirical results for $\lambda_k^{(l)}$ fit into the following combinatorial pattern: Let $\mathcal P(l)$ be the set of integer partitions of $l$. From $\mathcal P(l)$ we take only the set of partitions $P(l)$ with the property that any partition $p\in P(l)$ is given by a single even integer $g$, possibly zero, and $s$ odd integers\footnote{Here $s$ stands for any appropriate non-negative integer, that is of course not to be confused with the momentum parameter $s=1/t$.} $o_i$ with $1<o_1<o_2< \cdots < o_s$ such that
\begin{equation}
	l	=	g+\sum_{i=1}^s m_i o_i~,
\end{equation}
so $m_i$ is the multiplicity of $o_i$ in the respective partition of $l$. In this case we may write $p=( g,o_1^{m_i},\hdots , o_s^{m_s} )$. With this notation the combinatorial pattern of the coefficients $\lambda_k^{(l)}$, where $k=0,\hdots,l$, now reads%
 \footnote{With $\frac{(\pi i )^{2n}}{(2 n)!}=-\frac{\zeta(2 n)}{2^{2n-1} B_{2n}}$ we could also write the whole expression (\ref{lambda})  in terms of  zeta values and Bernoulli numbers.}%
\begin{equation} 
	\begin{array}{rl} 
		 \lambda_k^{(l)}	= \qquad &\displaystyle{(-1)^{k+1} \frac{(l+1)!}{k!} \,\,   \, \sum_{p\in P(l-k)} \ \  \frac{(-1)^\frac{g}{2}(\pi)^{g}}{ g!}  \prod_{i=1}^s \frac{2^{m_i}}{(o_i)^{m_i} \, m_i!} \zeta(o_i)^{m_i}} \\ 
                                  		 +  i \pi&\displaystyle{ (-1)^{k+1} \frac{(l+1)!}{k!} \,   \sum_{p\in P(l-k-1)}   \frac{(-1)^\frac{g}{2}(\pi)^{g}}{ (g+1)!}  \prod_{i=1}^s \frac{2^{m_i}}{(o_i)^{m_i} \, m_i!} \zeta(o_i)^{m_i}}~
.	\end{array}
\label{lambda}  
\end{equation} 
\noindent Indeed, we can give a generating function for the values $\lambda_0^{(l)}$ by
 \begin{gleichung}
 	\sum_{l=0}^\infty\lambda_0^{(l)}	\frac{x^l}{(l+1)!}	=	-\mathrm e^{i\pi x + \sum_{k=1}^\infty \frac{2\zeta(2k+1)}{2k+1}x^{2k+1}}	=	-\frac{\Gamma(1-x)}{\Gamma(1+x)}\mathrm e^{-2\gamma x+i\pi x}~,
 \label{generatinglambdas}
 \end{gleichung}
where $\gamma$ is the Euler-Mascheroni constant. The other coefficients are related to $\lambda_0^{(l)}$ by
\begin{equation}
	\lambda_k^{(l)}	=	(-1)^k \binom{l+1}{k}\lambda_0^{(l-k)}~.
\label{relationslambda}  
\end{equation}
Finally, we remark that $\sum_{k=0}^{l} \text{Im}(\lambda_k^{(l)}) \varpi_k$ is proportional to the vanishing period at $s=\tfrac{1}{(l+1)^2}$.

\subsection{The $\widehat \Gamma$-class and zeta values  at the point of maximal unipotent monodromy} 
\label{sec:gammaclass}

\noindent 
In this subsection we will  discuss  those aspects, in particular the importance of the number of moduli, of the Calabi-Yau geometries\footnote{We drop from now on the superscript expressing the number of moduli.} $M_{l-1}$  
shortly  introduced in subsection \ref{sec:l-loopbanadef}, which are most relevant to discuss the  
$\widehat \Gamma$-class formalism. The latter fixes the coefficients of the expansion of the Feynman 
amplitude \eqref{bananageneral} in terms of a canonical  Frobenius basis of solutions near the 
point of maximal unipotent monodromy. Eventually, this Frobenius basis can be related to period integrals  over an integral basis of 
cycles in the middle dimensional cohomology of the Calabi-Yau  geometry $M_{l-1}$ and a single chain integral extension.

Recall that upon numerical analytic continuation from the region  $1/s< (\sum_{i=1}^{l+1} \xi_i)^2$, where the 
amplitude can be calculated using  the Bessel function realization \eqref{besselform}, to the region 
$s<1/(\sum_{i=1}^{l+1} \xi_i)^2$ we got in the equal mass case and  for the first few loop orders $l\le 6$ the 
coefficients $\lambda_k^{(l)}$ displayed in Table \ref{tablambdaequal}.      
\begin{table}[htp]
{\footnotesize
\begin{center}\begin{tabular}{c?cccccc}
	$l $    &$ \varpi_0      $&$ \varpi_1 $&$\varpi_2$&$ \varpi_3 $&$ \varpi_4$ &$ \varpi_5$ \\[1ex]\toprule[1.5pt]
	$1 $  &$-2 \pi i           $&$                       $&$             $&$                $&$ $    &$ $ \\ [ 1mm]
	$2$   &$18\zeta(2)     $&$ 6 \pi i             $&$             $&$                $&$ $     &$ $ \\ [ 1mm]
         $3$   &$ -16 \zeta(3)+24 i \pi  \zeta(2)   $&$  - 72 \zeta(2)           $&$  -12 \pi i            $&$                $&$  $     &$ $ \\ [ 1mm]
         $4$   &$-450 \zeta (4)-80 i \pi  \zeta (3) $&$      80 \zeta(3)-120 \pi i \zeta(2)$&$ 180 \zeta(2)            $&$  20 \pi i               $&$  $     &$ $ \\ [ 1mm]
         $5$   &$ \begin{array}{c}
         -288 \zeta (5)+1440 \zeta (2) \zeta (3)- \\
                        540 i \pi  \zeta (4)\end{array}               $&$   \begin{array}{c}
      2700 \zeta (4)+  \\
          480 i \pi  \zeta (3)            \end{array}              $&$     \begin{array}{c}
       -240 \zeta(3)+ \\
            360 \pi i \zeta(2)          \end{array}             $&$  -360 \zeta(2)              $&$ -30 \pi i $     &$ $ \\ [ 1mm]
         $6$   &$\!\!\!\! \begin{array}{c}
                      6615 \zeta (6)-1120 \zeta (3)^2+\\
                      \pi i  (3360  \zeta (2) \zeta (3)-2016\zeta (5))\end{array}  \!\!\!\!                  $&$      \!\!\!\!         
                      \begin{array}{c}
                       2016 \zeta (5)-10080 \zeta (2) \zeta (3)+ \\
                     3780 i \pi  \zeta (4) \end{array}   \!\!             $&$    \!\!          
                       \begin{array}{c}
                     -9450 \zeta (4)- \\
                      1680 i \pi  \zeta (3)\end{array}      \!\!              $&$    \!\!\      
                      \begin{array}{c}
                     560 \zeta(3)- \\
                     840 \pi i \zeta(2) \end{array}        \!\!\!\!                  $&$  630 \zeta(2)   \!\!         $     &$  \!\!\         42 \pi i         $ 
\end{tabular}
\end{center}}
\caption{Numerically determined coefficients $\lambda^{(l)}_k$ for the  equal mass Feynman integral w.r.t. the Frobenius basis $\varpi_k$ at the MUM point for $l\le 6$. }
\label{tablambdaequal}
\end{table}%
Using further results up to $l=15$  we could guess the pattern summarized  in \eqref{lambda} or \eqref{generatinglambdas} combined with  \eqref{relationslambda}. 
This conjecturally determines all $l$-loop banana Feynman amplitudes  in all regions of their  physical  
parameter space, since the linear combinations for the non-equal mass case follow by a simple symmetric  splitting (see Table \ref{tablambdanonequal} below).
 
Generally, the occurrence of powers of zeta values and of some special numbers in Table~\ref{tablambdaequal}
that can be identified with Euler number integrals over combinations of top Chern classes, as well as equation \eqref{relationslambda}, 
suggest that the coefficients come from a $\widehat \Gamma$-expansion integrated against the exponential $e^{\mathfrak{t} \omega}$ of
a suitable  K\"ahler form $\omega$. Here we want to use the relevant Calabi-Yau varieties $M_{l-1}$ and $W_{l-1}$ together with a Fano geometry $F_l$ to prove the equations  
\eqref{lambda} or \eqref{generatinglambdas} using the $\widehat \Gamma$-class formalism.   

We start with the imaginary part of the numbers in  \eqref{lambda} or Table \ref{tablambdaequal}. The 
analytic continuation as well as the monodromy worked out in section \ref{sec:Monodromy} reveals that 
this combination of periods corresponds to the one that vanishes  at the nearest conifold 
$s=1/(\sum_{i=1}^{l+1} \xi_i)^2$. Geometrically here a sphere $S^{l-1}$ vanishes. The latter is in 
the integral middle cohomology of $M_{l-1}$ and by Seidel-Thomas twist it corresponds to the $D_{l-1}$ 
brane that wraps the full $(l-1)$-dimensional mirror Calabi-Yau space $W_{l-1}$ in the derived category of coherent sheaves on $W_{l-1}$.  

The $\widehat \Gamma$-class formalism\footnote{First explanations of the $\zeta(3)\chi/(2 (2 \pi i))^3$  and the $\int_W c_2\wedge\omega_k/24$ values  in the 
periods of three-folds as coming from derivates of the gamma function were made in~\cite{Hosono:1994ax}. See also \cite{MR1689204}.}~\cite{MR1838444,MR2683208,MR2483750,MR3536989}  
allows to calculate the  $K$-theory charge $Z_{D_{k}}$ of  
any D-brane  $D_k$ via
 \begin{equation}   
 Z_{D_k}(\mathfrak{t}) =\int_{W_{l-1}} e^{\omega\cdot \mathfrak{t}} \, \widehat \Gamma(TW_{l-1}){\rm  Ch}(D_k) +  {\cal O}(e^\mathfrak{t})\ .
\label{K-theorycharge} 
 \end{equation}    
 Here $\omega$ is the K\"ahler form of $W_{l-1}$ and $\omega  \cdot \mathfrak{t}=\sum_{i=1}^{h^{1,1}} \omega_i \mathfrak{t}^i$  
 refers to an expansion of the latter in terms of K\"ahler parameters $\mathfrak{t}^i$ w.r.t.  a basis  
 $\omega_i$  of the K\"ahler cone of $W_{l-1}$. ${\rm Ch}(D_k)$ defines a cohomology class that specifies the $D_k$ brane. 
 In particular, for the top dimensional $D_{l-1}$ brane ${\rm Ch}(D_{l-1})=1$. The 
 mirror map at the point of maximal unipotent monodromy\footnote{By  $\varpi^k_1$,  $k=1,\ldots, h^{l-2,1}(M_{l-1})$ we 
 denote all single logarithmic periods. If  $h^{l-2,1}(M_{l-1})=1$ we omit the upper index.}  
\begin{equation}
 \mathfrak{t}^k = \frac{1}{2 \pi i} \frac{\varpi^k_1}{\varpi_0}=\frac{1}{2 \pi i} \log (z_k)+ \tilde{\Sigma}_k(z) 
 \label{mirrormap}
 \end{equation} 
allows to relate the latter to the  corresponding period in the Frobenius basis. More precisely,  
the central charge $ Z_{D_{l-1}}$ is identified  with the period in question and the $\mathfrak{t}^k$-expansion 
can be identified with the logarithmic expansion in the Frobenius basis. In particular, to prove 
the occurence of the imaginary  terms in the first column of  Table  \ref{tablambdaequal}, we 
only need to expand the $\widehat \Gamma$-class of the tangent sheaf $TW_{l-1}$ of $W_{l-1}$. More generally, 
we consider the regularised  $\widehat \Gamma$-class of a sheaf ${\cal F}$. 
For a sheaf of rank $n$ the  latter is defined  as the symmetric expansion 
$\widehat \Gamma({\cal  F})=\prod_{i=1}^n e^{\gamma \delta_i} \Gamma(1+\delta_i)$ in terms of  the 
eigenvalues $\delta_i$ of  ${\cal F}$ which  in turn is re-expressed in terms of its Chern classes 
$c_k=s_k(\delta_1,\ldots, \delta_n)$.  Here $\gamma$ is  the Euler-Mascheroni constant and the $e^{\gamma \delta_i}$ factors are 
introduced to  cancel $e^{-\gamma c_1({\cal F})}$ terms that would arise from the first derivative of just the  $\Gamma(1+\delta_i)$ factors.   
For practical purposes it is faster to first write  a closed formula in terms of the Chern characters  ${\rm ch}_k({\cal F})$  of ${\cal F}$. 
That yields the regularized $\widehat \Gamma({\cal  F})$-class  as 
\begin{equation}
\widehat \Gamma({\cal F})=\exp\left(\sum_{k\ge 2}(-1)^k (k-1)!\, \zeta(k)\, {\rm ch}_k( {\cal F})\right) \ .    
 \end{equation}        
 The transition to the Chern classes  $c_k$  can be made by Newton's formula 
 \begin{equation}
 {\rm ch}_k=(-1)^{(k+1)} k\, \Bigg[ \log\left( 1+\sum_{i=1}^\infty c_i\, x^i\right) \Bigg]_k \ ,
\end{equation}    
  where $[\cdot]_k$ means to take the $k$th coefficient (in $x$) of the expansion of the expression in the $[\cdot]$-bracket.

 Let us now apply this to the geometry $W_{l-1}$ in (\ref{CICY}),  the Fano variety $F_l$ and its mirror $M_{l-1}$ using the mirror symmetry 
 formalism developed in~\cite{Hosono:1994ax}. Here  a canonical subfamily with $l+1$ complex structure deformations  of (\ref{CICY}) is 
 identified  with  the mirror manifold $M_{l-1}$  to $W_{l-1}$. We first want to establish that the Picard-Fuchs equations 
 and their solutions in the canonical Frobenius basis of $M_{l-1}$  are the same as the one that we derived for the 
 Feynman graph in the physical parametrization  (\ref{maxcutdef}). According to~\cite{Hosono:1994ax} the period solutions of complete 
 intersections in toric ambient spaces are specified  by $\ell$-vectors\footnote{The terminology of $\ell$-vectors employed here is of course not be confused with vectors 
 that have $l$ components or the like, which is why have chosen a different symbol 
 $\ell$.} $\ell^{(k)}=(\ell^{(k)}_{01},\ldots, \ell^{(k)}_{0h};  \ell^{(k)}_{1},\ldots , \ell^{(k)}_{c})$ for $k=1,\ldots, h^{l-2,1}(M_{l-1})$. 
 Here $h$ is the number of complete intersection constraints, $c$ is the number of homogenous coordinates of the ambient space  
 and the $\ell^{(k)}_{l}$ for $l=1,\ldots,c$, are the degrees of the constraints.   In the case of $M_{l-1}$ we have $h=2$ , $c=2(l+1)$ and the $\ell^{(k)}$ read
\begin{equation} 
\begin{array}{rl} 
\ell^{(1)}=&(-1,-1;1,1,0,0,\cdots ,0,0,0,0)\\
\ell^{(2)}=&(-1,-1;0,0,1,1,\cdots ,0,0,0,0)\\
\vdots \quad    \phantom{=}&    \\
\ell^{(l)}=&(-1,-1;0,0,0,0,\cdots ,1,1,0,0)\\
\ell^{(l+1)}=&(-1,-1;0,0,0,0,\cdots ,0,0,1,1)\ \, .
\end{array} 
\label{ell} 
\end{equation}
From these   $\ell$-vectors one obtains a generalized  Gelfand-Kapranov-Zelevinsk\u{\i} differential system with holomorphic solution  
\begin{equation} 
\omega_0({\underline z};{\underline \epsilon})=\sum_{n_1, \dots, n_{l+1}\ge 0 } c({\underline{n}+\underline{\epsilon}}) \, {\underline z}^{\underline n+\epsilon}\, .
\label{holsolwithyau} 
\end{equation}  
Here the underlined  quantities  are $(l+1)$-tuples and the series coefficients $c(\underline{n})$ are determined by the $l+1$ $\ell$-vectors via
\begin{equation} 
c(\underline{n})=\frac{
\prod_{j=1}^2 \left(-\sum_{k=1}^{l+1} l_{0j}^{(k)} \, n_k \right)!}
{\prod_{i=1}^{2l+2}\left(\sum_{k=1}^{l+1} l_i^{(k)}\, n_k \right)!} \, .
\label{defcn}
\end{equation} 
The $c(\underline{n}+\underline{\epsilon})$  are as usual defined by promoting all factorials $*!$ in (\ref{defcn}) to $\Gamma(*+1)$ and 
deforming each integer  $n_k$ to  $n_k+\epsilon_k$. In particular, the unique holomorphic solution at the point of maximal 
unipotent monodromy is given by
\begin{equation} 
\varpi_0({\underline z})=\omega_0({\underline z};{\underline \epsilon })|_{{\underline \epsilon }={\underline 0 }}= \sum_{n_1, \dots, n_{l+1}\ge 0 } \binom{|n|}{n_1,\hdots,n_{l+1}}^2 \prod_{k=1}^{l+1} z_k^{n_k}	\, .
\label{defw0}
\end{equation} 
Comparing with \eqref{maxcutmum} we see that the coordinates\footnote{These coordinates are also related to 
the redundant parameters multiplying the generic monomials of the Newton polytope associated to the complete intersection Calabi-Yau, see~\cite{Hosono:1994ax} for a definition.} $z_k$ are related to the physical coordinates by
\begin{equation}
	z_k	=	s \xi_k^2	\quad\text{for } k=1,\hdots,l+1~.
\label{redf}
\end{equation}
The point  here is that the period of $M_{l-1}$ given in \eqref{defw0} is up to the simple parameter redefinition \eqref{redf} equivalent 
to \eqref{maxcutmum} after multiplying with the physical variable $s$.  The other periods for both systems can be obtained by the Frobenius 
method as described in section \ref{ssc:frobeniusEM} for the Feynman graph period and in~\cite{Hosono:1994ax} for the periods of $M_{l-1}$. The basic 
idea  is to take certain combinations of derivatives w.r.t the deformations parameters $L^{r}_{c}=\sum_{j_1,\ldots, j_r} c_{j_1,\ldots,  j_r} 
\partial_{\epsilon_{j_1}} \ldots  \partial_{\epsilon_{j_r}}   \omega_0({\underline z};{\underline \epsilon})|_{{\underline \epsilon }={\underline 0 }}$. 
In particular, the  $\varpi_k=L^{(1)}_{\delta_{k,j}}/2 \pi i$, $k=1,\ldots, h^{l-2,1}(M_{l-1})=l+1$  are the  single logarithmic solutions, which  
 together with $\varpi_0$ determine the mirror map (\ref{mirrormap}).  The higher  logarithmic solutions are fixed by the 
 topological data of $W_{l-1}$ and their numbers inferred by the differential ideal  reported  in Table  \ref{tablelogsolution}
 matches the primitive vertical  Hodge numbers of $W_{l-1}$  and primitive horizontal middle dimensional Hodge numbers of $M_{l-1}$ 
 discussed in~\cite{Klemm:1996ts,Bizet:2014uua} for four-folds.
  
These identifications suggest that  \eqref{CICY} is the right mirror $W_{l-1}$ to the Calabi-Yau ($l-1$)-fold family $M_{l-1}$  whose periods, together with the single chain integral extension, in the parametrization in \eqref{redf} describe non-redundantly the  Feyman graphs exactly in the physical parameters.         
 
 As we explained in the beginning of this subsection this  implies that the evaluation of the $\widehat \Gamma$-class for $W_{l-1}$ must 
 reproduce the  imaginary parts of  (\ref{lambda}). It follows by the adjunction formula that the Chern classes $c_k$ of $W_{l-1}$ are given by
 the degree $k$ part of 
 \begin{equation}
 c_k(W_{l-1})=\left[\frac{\prod_{i=1}^{l+1}(1+H_i)^2}{(1+ \sum_{i=1}^{l+1} H_i)^2}\right]_{\text{deg}(H)=k}\ .
 \label{c_k} 
 \end{equation}     
More precisely, since the hyperplane classes in each $\mathbb{P}^1$  fulfill $H_i^2=0$ we can 
express $c_k$ in terms of elementary symmetric polynomials $s_k({\underline H})=\sum_{i_1<\ldots < i_k} H_{i_1}\cdots H_{i_k}$ 
as 
\begin{equation} 
c_k(W_{l-1})=(-1)^k k! \sum_{j=0}^k \frac{(-2)^j  (k+1-j)}{j!} s_k({\underline H})=:{\cal N}^{W_{l-1}}_k s_k({\underline H})   \ . 
\end{equation}    
Similarly, considering the power one of the normal bundle in the denominator of (\ref{c_k}) (instead of two) we can write   for the Chern classes of the ambient space
\begin{equation} 
c_k(F_l)=(-1)^k k! \sum_{j=0}^k \frac{(-2)^{j}}{j!} s_k({\underline H})=:{\cal N}^{F_l}_k s_k({\underline H})   \ . 
\end{equation}   
Moreover, we notice that the integral of a top degree product of  Chern classes $c_{k_n}$ over $X=W_{l-1}$ or $X=F_l$ is
given  by
\begin{equation}
\int_{X} \prod_n c_{k_n}= (l+1)!\, \prod_{n} \frac{{\cal N}^X_{k_n}}{k_n!} \ .
\end{equation}  

Let us comment some more on primitive vertical homology of $W_{l-1}$ and well as $F_l$ and establish that  $W_{l-1}$ 
is the mirror  of $M_{l-1}$ for the three-fold case. Let us denote the homogenous coordinates of the 
$k$th $\mathbb{P}^1_k$ by $[x_k:y_k]$. Then in general the independent divisor classes of  
$W_{l-1}$ and as well as of $F_l$ are the restrictions of the hyperplane classes\footnote{To ease the notation we  denote the divisor classes on 
$W_{l-1}$ and $F_l$ again by $D_k$.}  $D_k=\{a x_k+ by_k=0\} $ in $(\mathbb{P}_1)^{l+1}$ with topology $(\mathbb{P}_1)^{l}$ for 
$F_l$ or $W_{l-1}$, where they have  topology $F_{l-1}$ or $W_{l-2}$, respectively.  We have $D_k^2=0$, $k=1,\ldots,l+1$, and the top 
intersections are encoded in the coefficients of the rings
\begin{equation} 
{\cal R}^{W_{l-1}}=2 \sum_{i_1< \ldots < i_{l-1}} D_{i_1}\ldots D_{i_{l-1}} \quad\text{and}\quad  {\cal R}^{F_{l}} =\sum_{i_1< \ldots <i_{l}} D_{i_1}\ldots D_{i_{l}}\ .
\label{Intersectionring} 
\end{equation}   
The $D_k$ generate the primitive part of the vertical cohomology of $(\mathbb{P}^1)^{l+1}$, for which it is the full vertical cohomology with 
dimensions $h^{k,k}\left(\left(\mathbb{P}^1\right)^{l+1}\right)=\left(l+1\atop k\right)$, as well as for $F_l$ and $W_{l-1}$, for which it is 
\begin{equation}
h_{\rm prim} ^{k,k}(F_{l})=\left\{ \begin{array}{rl} 
\left(l+1\atop k\right) & {\rm if }\ k< \left\lceil \frac{l}{2}\right\rceil \\
\left(l+1\atop l-k\right) & {\rm otherwise }
\end{array}\right.
\quad\text{and}\quad
 h_{\rm prim}^{k,k}(W_{l-1})=\left\{ \begin{array}{rl} 
\left(l+1\atop k\right) & {\rm if }\ k< \left\lceil \frac{l}{2}\right\rceil-1 \\
\left(l+1\atop l-1-k\right) & {\rm otherwise }
\end{array}\right.~.
\label{primcohom}
\end{equation} 
For high dimensions the primitive part is much smaller than the full vertical cohomology. The latter fact can be easily seen by calculating via 
the Hirzebruch-Riemann-Roch  index theorem  the elliptic genera $\chi_k=\sum (-1)^q h^{q,k}(X)$  by evaluating instead of 
the $\widehat \Gamma$-class the Todd class against $\text{Ch}(\wedge^k TX)$.    

According to~\cite{Klemm:1996ts,Bizet:2014uua}  this primitive part of the vertical cohomology  should be mirror dual  to the primitive horizontal  middle cohomology
which corresponds  to solutions of those Picard-Fuchs equations as discussed in subsection \ref{ssc:non-eq-mass-diff}. Luckily, it is  only those  
solutions we  need  to describe the banana diagrams. The actual vertical-- and horizontal cohomology groups are much bigger. For example for 
the differently polarized K3 surfaces called $M_2$ and $W_2$ we have $h^{1,1}_{\rm vert\ prim}(W_2)= h^{1,1}_{\rm hor\ prim}(M_2)= 4$ inside the twenty-dimensional group $H_{1,1}(\text{K3})$.

The $D_k$  are dual to the rational curves $C_k=\mathbb{P}^1_k$ which span the Mori cone. The latter pair by integration $\int_{C_j}\omega_i=\delta_{ij}$ 
with the  K\"ahler forms $\omega_k$  of the $\mathbb{P}^1_k$, which span the K\"ahler cone. To see e.g. that $M_3$ is really the 
mirror to $W_3$ we can check the mirror symmetry predictions at the level of the instantons. Using  equations \eqref{ell} and \eqref{Intersectionring} we 
apply\footnote{This can be done with the program \texttt{Instanton} distributed with~\cite{Hosono:1994ax}.}   the formalism of~\cite{Hosono:1994ax}  for the prepotential of the case at hand and get up to order ${\cal O}(q^{11})$ and up to permutations 
$$
\begin{array}{rl} 
{\cal F}^{\rm prep}
&=\displaystyle{ 2 \sum_{i< j < k }\mathfrak t^i \mathfrak t^j \mathfrak t^k  + \sum_{i=k}^5 \frac{ 24}{24} \mathfrak t^k -80\frac{ \zeta(3)}{2 (2 \pi i)^3}+ 24 \, {\rm Li }_3(q_1)+ 24 \, {\rm Li }_3(q_1q_2) +
112 \, {\rm Li }_3(q_1 q_2 q_3)} 
\\ &  \quad
+\num{1104} \, {\rm Li }_3(q_1 q_2q_3q_4) + \num{19200} \, {\rm Li }_3(q_1 q_2 q_3 q_4 q_5) + 24 \, {\rm Li }_3(q_1^2 q_2 q_3) + \num{1104} \, {\rm Li }_3(q_1^2 q_2 q_3 q_4) 
\\ &  \quad
+\num{45408} \, {\rm Li }_3(q_1^2 q_3 q_4 q_5) +24 \, {\rm Li }_3(q_1^2 q_2^2 q_3) + \num{2800} \, {\rm Li }_3(q_1^2 q_2^2 q_3 q_4) +
 \\ &\quad
 + \num{212880} \, {\rm Li }_3(q_1^2 q_2^2 q_3 q_4 q_5) + 80 \, {\rm Li }_3(q_1^2q_2^2q_3^2) +   \num{14496} \, {\rm Li }_3(q_1^2q_2^2q_3^2 q_4)\!  
 \\&\quad
 +\num{1691856} \, {\rm Li }_3(q_1^2q_2^2q_3^2 q_4 q_5) + \num{122352} \, {\rm Li }_3(q_1^2q_2^2q_3^2 q_4)\ . \\
\end{array}
$$
Here the $q_k=\exp(2 \pi i \mathfrak t^k)$ keeps track of the (multi-) degree of a rational  instanton contribution. 
${\rm Li}_k(x)=\sum_{n=1} \frac{x^n}{n^k}$ denotes the polylogarithm 
and ${\rm Li}_3(x)$  subtracts the multi covering contributions to the $g=0$ curves.  
The integral coefficients   of ${\rm Li}_3({\underline q}^{\underline d})$ are denoted by  $n_0^{\underline d}$. If 
the curves are smooth $n_0^{\underline d}=(-1)^{{\rm dim}({\cal M}_{\underline d})} e({\cal M}_{\underline d})$ is up to sign the Euler number of the 
moduli space ${\cal M}_{\underline d}$  of the rational curves of degree ${\underline d}$.  We see that these instanton numbers  
are indeed as expected for $W_3$. For example each single degree one curve  gives a contribution as $24 \, {\rm Li }_3(q_1)$. Since 
the  moduli space of such a curve is the K3 over which the $\mathbb{P}^1$ is fibered  we get indeed $n^{(1,0,0,0,0)}_0= (-1)^2 \chi(\text{K3})=24$.  
The  geometry $W_{l-1}$ has an intriguing nested fibration structure. For example  the K3 called $W_2$ is in four ways fibred by the elliptic 
curve $W_1$  over $\mathbb{P}_k^1$ for $k=1,\ldots,4$. While the Calabi-Yau three-fold $W_3$ for $l=4$ is in five ways fibered\footnote{The latter fact 
can be checked  using the criterium of Oguiso~\cite{MR1228584} from the topological data that appears in the classical part of ${\cal F}^{ \rm prep}$.} with a 
K3  fiber of topology $W_2$, etc.

Now we can come to the main point and can use the mirror picture, the $\widehat\Gamma$-class conjecture and evaluation of the  
Chern classes to show that the leading logarithms (or $\mathfrak{t}$ powers) in the  evaluation  of the $\widehat \Gamma$-class 
\begin{equation}   
 Z_{D_{l-1}}(\mathfrak{t}) =\int_{W_{l-1}} e^{\omega\cdot \mathfrak{t}} \, \widehat \Gamma(TW_{l-1}) +  {\cal O}(e^\mathfrak{t})\ 
\label{highestCYK-theorycharge} 
\end{equation}
yield precisely the imaginary parts of Table \ref{tablambdaequal}  or more generally  of  (\ref{lambda}).

Furthermore, we checked\footnote{A few days after this preprint appeared in the arXiv, we received detailed e-mails from Hiroshi Iritani~\cite{privcomiritani} indicating that the formula can be proven using the formalism~\cite{MR2683208,MR3536989}.}  that the leading   logarithms (or $\mathfrak{t}$ powers) in the evaluation of 
\begin{equation}   
 Z_{{\cal F}_{\sigma_l}}(\mathfrak{t}) =%
 			\int_{F_l} e^{\omega\cdot \mathfrak{t}} \, \frac{1}{\widehat \Gamma^2(TF_l)}\widehat A(TF_l) +  {\cal O}(e^\mathfrak{t}) 
			=	\int_{F_l} e^{\omega\cdot \mathfrak{t}} \, \frac{\Gamma(1-c_1)}{\Gamma(1+c_1)}\cos(\pi c_1)	 +  {\cal O}(e^\mathfrak{t})	
\label{K-theorychain} 
\end{equation}
yield precisely the real part of Table \ref{tablambdaequal}  or more generally  of  (\ref{lambda}). Here $\widehat A(TF_l)$ is the $\widehat A$-genus of the tangent sheaf $TF_l$ which is generally defined for a rank $n$ sheaf $\mathcal F$ as the symmetric expansion $\widehat A(\mathcal F)=\prod_{i=1}^n\frac{2\pi i \delta_i}{\sinh(2\pi i\delta_i)}$ in terms of the eigenvalues $\delta_i$ of $\mathcal F$ and rewritten in terms of the Chern classes $c_k$.

Some remarks about \eqref{K-theorychain} have to be said. First of all the two equalities in \eqref{K-theorychain} were found by fitting an ansatz of a generalization of the $\widehat\Gamma$-class to the analytically computed values in \eqref{lambda}. By this we observed that ansatze as in \eqref{K-theorychain} reproduce the correct $\lambda$-coefficients. Notice that we are pretty sure that our ansatze are very special to the Fano variety $F_l$ we are considering here. We do not believe that the relations in \eqref{K-theorychain} are generally true. We leave it to future work to geometrically interpret and prove our modified $\widehat\Gamma$-class conjecture, see~\cite{Matt:progress}. Secondly, we remark that the expected usual $\widehat\Gamma$-class ansatz as in \eqref{highestCYK-theorycharge} does not work even when ambiguities in the $\lambda$-coefficients are taken into account. These ambiguities arise due to analytic continuation of the Feynman amplitude around the MUM point. The monodromy of the Feynman amplitude is given by shifting each logarithm by $2\pi i$ such that $\lambda$-coefficients which contain even zeta values are shifted or in other words are ambiguous. Due to this ambiguity contributions containing only odd zeta values are ambiguity free and can naively be matched to the $\widehat\Gamma$-class. But also these terms are not correctly reproduced by a simple $\widehat\Gamma$-class\footnote{For example for $l=9$ the contribution of $\zeta(3)^3$ or for $l=11$ the contribution of $\zeta(5)\zeta(3)^2$.}. Therefore, we take into account the more general form as given in \eqref{K-theorychain} fitting to all even and odd zeta values in the $\lambda$-coefficients. Moreover, notice that the second equality in \eqref{K-theorychain} is very useful since from it the real part of \eqref{relationslambda} follows trivially because the integral over $F_l$ yields simply a contribution of $(l+1)!$ for $c_1^l$. But this term is again canceled in the generating series \eqref{relationslambda}.



 \subsection{Monodromy}
 \label{sec:Monodromy}
To each singular point $s'$ of $\mathcal{L}_l$ (recall equation \eqref{discriminantequal}) we can associate a monodromy matrix $M^{(l)}_{s'}$ acting on the Frobenius basis $\varpi_0,\hdots,\varpi_l$ around the MUM point $s=0$ by
\begin{align}
    \left( \begin{array}{c}
         \varpi_0 \\
         \vdots\\
         \varpi_l
    \end{array} \right) \longmapsto M^{(l)}_{s'} \left( \begin{array}{c}
         \varpi_0 \\
         \vdots\\
         \varpi_l
    \end{array} \right)~,
\end{align}
where we choose the analytic continuation along the upper half plane and encircle the singular point $s'$ counterclockwise. For the MUM point $s'=0$ one can directly read off $M^{(l)}_0$ from the structure of the Frobenius basis. At the singular points 
\begin{align}
     \frac{1}{(l+1)^2}~, \, ...~, \, \frac{1}{(l+1-2\lfloor \frac{l+1}{2} \rfloor)^2}
\end{align}
the local Frobenius basis can in each case be chosen so that only one solution is singular, i.e. for these points the monodromy satisifies
\begin{align}
    \dim \left(\text{image} (M^{(l)}_{s'}-\mathds{1})\right)	=	1~.
    \label{eq:DimMon}
\end{align}
This motivates the definition of the Frobenius constants $\kappa^{(l,s')}_k$ by
\begin{align}\label{eq:defkappas}
    (M_{s'}^{(l)}-\mathds{1})\varpi_k=\kappa^{(l,{s'})}_k(M_{s'}^{(l)}-\mathds{1})\varpi_0~,
\end{align}
i.e.\ we choose the normalization such that $\kappa_0^{(l,{s'})}=1$. Note that this only works when $\varpi_0$ is not invariant under $M_{s'}^{(l)}$.  

The Feynman amplitude $\mathcal{F}_{\sigma_l}$ is only singular at $s'=0$ and $s'=\tfrac{1}{(l+1)^2}$. For all other singular points $s'$ satsifying (\ref{eq:DimMon}) this implies\footnote{Insert the definitions \eqref{eq:deflambda} and \eqref{eq:defkappas} in the trivial monodromy condition $(M_{s'}^{(l)}-\mathds{1})\mathcal{F}_{\sigma_l}=0$. } that the Frobenius constants are constrained by
\begin{align}
    \sum_{k=0}^l \lambda^{(l)}_k \kappa^{(l,s')}_k=0~.
\end{align}
This can not hold for $s'=\tfrac{1}{(l+1)^2}$. However, numerically we find even stronger conditions for this point, i.e.\ we observe that the Frobenius constants do not depend on the loop order,
\begin{align}
    \kappa_k^{(l,1/(l+1)^2)}=\kappa_k^{(l',1/(l'+1)^2)}\quad \mathrm{for} \  k \leq l,l'~,
\end{align}
and that
\begin{align}
    \sum_{k=0}^l\kappa_k^{(l,1/(l+1)^2)}\lambda_k^{(l)}=-(2\pi i)^l~.
\end{align}
Thus, restricting to the singular point $s'=\tfrac{1}{(l+1)^2}$, there is a series $\kappa_k$ of Frobenius constants determined by
\begin{align}
    \sum_{k=0}^\infty \frac{\kappa_k}{k!}x^k	=	\frac{1}{\Gamma(1+x)^2}~ \mathrm e^{-2\gamma x} ~.
\end{align}
In terms of these Frobenius constants we can write the associated monodromy matrix as
\begin{align}
    M^{(l)}_{1/(l+1)^2}=\mathds{1}+ \left(
\begin{array}{c}
 \kappa_0 \\
 \vdots \\
 \kappa_l
\end{array}
\right) \left(\delta^{(l)}_0,..., \delta^{(l)}_l\right)
\end{align}
for some constants $\delta^{(l)}_k$. These constants can now be determined using the fact\footnote{The combined path of analytic continuation corresponding to the left hand side of \eqref{eq:monodromyid} is contractible in $\mathbb{P}^1$, as $\mathcal{F}_{\sigma_l}$ picks up no non-trivial monodromies at the other singular points.}
\begin{align}\label{eq:monodromyid}
    M^{(l)}_0 M^{(l)}_{1/(l+1)^2}\mathcal{F}_{\sigma_l}=\mathcal{F}_{\sigma_l}~,
\end{align}
which gives
\begin{align}
    \sum_{l=1}^\infty \frac{\delta^{(l)}_0}{(l+1)!}x^l	=	-\frac{x}{\Gamma(1-\frac x{2\pi i})^2}~\mathrm e^{\gamma \frac{x}{i\pi}}	
\end{align}
and the relation
\begin{align}
    \delta^{(l)}_k=\frac{1}{(2\pi i)^k}\binom{l+1}{k}\delta^{(l-k)}_0~.
\end{align}

\subsection{Special Points}
\label{subsec:specialpoints}
Special Feynman integrals can evaluate to interesting values related to critical L-values. Many relations of this type were conjectured by Broadhurst in \cite{Broadhurst:2016myo} and some also have been proven later. To summarize some of these we define the Hecke eigenforms
\begin{alignat}{2}
    f_3(\tau)&=(\eta(3\tau)\eta(5\tau))^3+(\eta(\tau)\eta(15\tau))^3 &&\in S_3\left(\Gamma_0(15),\left( \frac{-15}{\cdot}\right)\right)^{\text{new}}\\
    f_4(\tau)&=(\eta(\tau)\eta(2\tau)\eta(3\tau)\eta(6\tau))^2 &&\in S_4(\Gamma_0(6))^{\text{new}}\\
    f_6(\tau)&=\left( \frac{\eta(2\tau)^3\eta(3\tau)^3}{\eta(\tau)\eta(6\tau)}\right)^3+\left( \frac{\eta(\tau)^3\eta(6\tau)^3}{\eta(2\tau)\eta(3\tau)}\right)^3 &&\in S_6(\Gamma_0(6))^{\text{new}}
\end{alignat}
in terms of the Dedekind eta function $\eta$. For the equal mass Banana diagrams there are then the following proven relations for the on-shell point $s=1$ and loop orders $l\leq 6$:

\begin{table}[H]
\centering
{\small{
\begin{tabular}{l cp{12cm}@{}}
\toprule
Relation & Proof \\ \midrule
$\mathcal{F}_{\sigma_1}(1)=2L(\left( \frac{\cdot}{3}\right),1)=\frac{2\pi}{3^{3/2}}$ & Bailey et al.: \cite{bailey2008elliptic} \\[1ex]
$\mathcal{F}_{\sigma_2}(1)=(1-2^{-2})\zeta(2)=\frac{\pi^2}{4}$ & Bailey et al.: \cite{bailey2008elliptic}\\[1ex]
$\mathcal{F}_{\sigma_3}(1)=\frac{12 \pi}{\sqrt{15}}L(f_3,2)=\frac{1}{30\sqrt{5}}\prod_{k=0}^3\Gamma \left( \frac{2^k}{15}\right)$ & Zhou: \cite{Zhou:2017vhw}\\[1ex]
$\mathcal{F}_{\sigma_4}(1)=8 \pi^2 L(f_4,2)$ &Zhou: \cite{Zhou:2017vhw}\\[1ex]
$\mathcal{F}_{\sigma_6}(1)=144\pi^2 L(f_6,4)$ &Zhou:  \cite{Zhou:2017vhw}\\[1ex]
\bottomrule
\end{tabular}
}}
\caption{Special values of the equal mass Feynman amplitudes at $t=1$}
\label{SpecialValues}
\end{table}

%
As was explained to us by Matt Kerr\footnote{Private communication}, since the banana Feynman integral is always a higher normal function (following from arguments similar to those in \cite{Bloch:2014qca}) and its Milnor symbol becomes torsion at $t=1$ for any loop order $l$, the integral evaluates (up to factors of $2\pi i$) to a period of the underlying Calabi-Yau manifold. For modular Calabi-Yau manifolds it is expected that these periods evaluate to special $L$-function values of the associated modular forms. For instance, $f_3$ is associated with a K3 surface of \cite{MR1184764} and $f_4$ is associated with a rigid Calabi-Yau threefold \cite{hulek2005modularity}.

Besides rigid Calabi-Yau three-folds defined over $\mathbb{Q}$, there are other interesting special Calabi-Yau threefolds $X$ associated with the four-loop diagram and suitable values of the kinematic parameters. One example are rank two attractor varieties. These are defined by demanding that their rational middle cohomology splits to 
\begin{align}
    H^3(X,\mathbb{Q})=\Lambda \oplus \Lambda_\perp
\end{align}
with 
\begin{align}
    \Lambda \subset H^{(3,0)}(X) \oplus H^{(0,3)}(X) \qquad \text{and} \qquad \Lambda_\perp \subset H^{(2,1)}(X) \oplus H^{(1,2)}(X)~.
\end{align}
For the four-loop Banana diagram, Candelas et al. found in~\cite{Candelas:2019llw} that the points $s=-1/7$ and $s=33\pm 8 \sqrt{17}$ correspond to rank two attractor varieties. For all of these points the Hasse-Weil zeta function gives rise to modular forms of weight two and four and in \cite{BoenischThesis} it was established numerically that the period matrices can be completely expressed in terms of periods and quasiperiods of these modular forms. E.g.\ for $s=-1/7$ one finds that the associated period matrix $T_{-1/7}$ can be written as
\begin{align*}
    T_{-1/7}=A\left(
\begin{array}{cccc}
 \omega_4^+ & \eta_4^+ & 0 & 0 \\
 \omega_4^- & \eta_4^- & 0 & 0 \\
 0 & 0 & \tilde{\omega}_2^+ & \tilde{\eta}_2^+ \\
 0 & 0 & \tilde{\omega}_2^- & \tilde{\eta}_2^- 
\end{array}
\right)B~,
\end{align*}
where $A$ and $B$ are matrices with rational entries, $\omega_4^\pm$ and $\eta_4^\pm$ are periods and quasiperiods of a modular form of weight four and $\tilde{\omega}_2^\pm$ and $\tilde{\eta}_2^\pm$ are (up to a factor of $2\pi i$) periods and quasiperiods of a modular form of weight two. The same is true for the other two points $s=33\pm 8\sqrt{17}$. However, in this case one has to take linear combinations over $\mathbb{Q}[\sqrt{-1},\sqrt{-2},\sqrt{17}]$ due to the coefficient fields of the modular forms and the irrationality of $s$, as explained in more detail in~\cite{BoenischThesis}. 

For the Feynman integral this means that the maximal cut integral at these points can be expressed in terms of the associated periods and quasiperiods. However, we found that this does not hold for the complete Feynman integral and it would be interesting to find out if and how this is related to the modularity of the underlying Calabi-Yau threefold.

\section{The $l$-loop non-equal mass banana Feynman integrals}
\label{sec:nonequalmass}
Having discussed the equal mass banana Feynman graph, we want to focus in this section on the full non-equal mass case. We give a general description and method how to compute the $l$-loop non-equal mass banana Feynman graph, exemplified by the four-loop non-equal mass case. In \cite{Klemm:2019dbm} we have already discussed the two- and three-loop case.

\subsection{Batyrev coordinates and the maximal cut integral}
As in \cite{Klemm:2019dbm} we use Batyrev coordinates\footnote{Notice that these are not the same Batyrev coordinates as defined in \eqref{redf} which correspond to the complete intersection model. Here now we consider the hypersurface model with other Batyrev coordinates.} $z_k$, defined by
\begin{gleichung}
	z_k	=	\frac{\xi_k^2}{t-\sum_{i=1}^{l+1}\xi_i^2}	\qquad\text{for } k=1,\hdots,l+1~.
\label{batyrevcoords}
\end{gleichung}
Furthermore, we include in the Feynman integral $\mathcal F_{\sigma_l}(t,\xi_i)$ the additional factor
\begin{equation}\label{eq:a0}
a_0	=	t-\sum_{i=1}^{l+1}\xi_i^2	=	\frac{\xi_{l+1}^2}{z_{l+1}}~,
\end{equation}	
which is related to the inner point of the polytope described by the polynomial constraint $P_l(t,\xi_i;x)=0$. Then the expression we want to determine is 
\begin{gleichung}
	\hat{\mathcal F}_{\sigma_l}	=	\int_{\sigma_l}	\frac{a_0 \mu_l}{P_l(t,\xi_i;x)}~,
\label{feynmanwithinnerpoint}
\end{gleichung}
and will $\hat{\mathcal F}_{T^l}$ be defined by analogy.
For large momenta we can use the expression \eqref{maxcutmum} to find the non-equal mass maximal cut Feynman integral\footnote{Again up to normalization.} including the inner point
\begin{gleichung}
	\hat\varpi_0(z_i)	=	\sum_{ k_1, \dots , k_{l+1} \geq 0 } \binom{|k|}{k_1,\hdots,k_{l+1}}^2\left(\frac{1}{1+\sum_{i=1}^{l+1}z_i}\right)^{1+|k|}\prod_{i=1}^{l+1}z_i^{k_i}~.
\label{maxcutmumnonequal} 
 \end{gleichung}
 Geometric series expansion gives a power series in the $z_i$  with non-negative exponents, valid if all $z_k$ are sufficiently small. The radius of convergence can be determined by the discriminant of the polynomial constraint $P_l(t,\xi;x)$ or later also from the differential operators annihilating \eqref{maxcutmumnonequal}. We claim that the discriminant\footnote{Or at least the discriminant factors up to multiplicities.} for the generic mass banana Feynman graph is given by
\begin{equation}
	\Delta(\mathcal D^l)(t,\xi_i)	=	
					t\prod_{ \{T_1,T_2\} \in \mathcal T}\left[ t-\left( \sum_{i\in T_1}\xi_i-\sum_{i\in T_2}\xi_i\right)^2\right]~,
\label{discriminantnonequal}
\end{equation}
where the set
\begin{equation}
\mathcal T= \Big\lbrace \{T_1, T_2 \} \, \Big| \, T_1, T_2 \subseteq T^{(l+1)} \, \text{disjoint} \ \, \text{and} \, \ T_1 \cup T_2 = T^{(l+1)} \Big\rbrace 
\end{equation}
gives all possibilities to distribute the $l+1$ indices among two subsets (identifying swaps of the two sets).
%
We have explicitly checked $\Delta(\mathcal D^l)(t,\xi_i)$ for $l=2,3,4$. The discriminant in the four-loop case is for instance given by
\begin{gleichung}
	\Delta(\mathcal D^4)(t,\xi_i)	=	t&	\left(t-(-\xi_1+\xi_2+\xi_3+\xi_4)^2\right)	\left(	t-(+\xi_1-\xi_2+\xi_3+\xi_4)^2\right)	\\
					&	\left(t-(+\xi_1+\xi_2-\xi_3+\xi_4)^2\right)	\left(	t-(+\xi_1+\xi_2+\xi_3-\xi_4)^2\right)	\\
					&	\left(t-(-\xi_1-\xi_2+\xi_3+\xi_4)^2\right)	\left(t-(-\xi_1+\xi_2-\xi_3+\xi_4)^2\right)	\\
					&	\left(t-(-\xi_1+\xi_2+\xi_3-\xi_4)^2\right)	\left(t-(+\xi_1+\xi_2+\xi_3+\xi_4)^2\right)~.
\end{gleichung}
In the equal mass case the discriminant $\Delta(\mathcal D^l)(t,\xi_i)$ reproduces the correct factors as stated in \eqref{discriminantequal}.

\subsection{Differential equations for the non-equal mass case} 
\label{ssc:non-eq-mass-diff}
Having found the holomorphic power series \eqref{maxcutmumnonequal} describing the maximal cut Feynman integral for small values of $z_k$, as in \cite{Klemm:2019dbm} we now want to find a set of (in)homogeneous differential equations for it. With the help of \eqref{maxcutmumnonequal} 
 it can be checked that the second order operators\footnote{At this point we want to mention that it is also possible the obtain differential operators from the complete intersection model in \ref{sec:l-loopbanadef}. They follow directly from the $\ell$-vector description, see~\cite{Hosono:1994ax}, and are closely related to the ones in \eqref{homops}. The following discussion could have similarly been made also starting from these operators and analyzing the complete intersection model. Nevertheless, we focus in our discussion on the hypersurface model since this is how we originally invented our method.} 
\begin{gleichung}
	\mathcal D_k	=	\theta_k^2		-	z_k\left(\sum_{i=1}^{l+1}\theta_i-2\theta_k\right)\left(1+\sum_{i=1}^{l+1}\theta_i\right)-z_k\left(\sum_{i=1}^{l+1}z_i-z_k\right)\left(1+\sum_{i=1}^{l+1}\theta_i\right)\left(1+\sum_{i=1}^{l+1}\theta_i\right)
\label{homops}
\end{gleichung}
for $k=1,\hdots,l+1$ annihilate $\hat\varpi_0(z_i)$. Applying these operators to the full Feynman integral $\hat{\mathcal F}_{\sigma_l}$ and performing a numerical integration we find that the operators $\mathcal D_k$ are indeed homogeneous operators annihilating the full Feynman integral $\hat{\mathcal F}_{\sigma_l}$, see also section 2.3 in \cite{Klemm:2019dbm}.

It turns out that these operators are enough to determine all solutions needed for the Feynman integral $\hat{\mathcal F}_{\sigma_l}$ --- including those from integrals over closed cycles as well as the additional solution arising due to the chain integral ---  once the correct structure of solutions is imposed. Recall that the $z_i$ are local coordinates around a MUM-point ($z_i=0$ for all $i=1,\dots, l+1$), so there is a unique holomorphic solution up to normalization and the rest of the local Frobenius basis is spanned by solutions with increasing degree of the leading logarithms.\footnote{Here the notion of degree or better multidegree is such that for instance the periods $\hat{\varpi}_k^r$ in eqs. \eqref{loglow} have degree $r$, i.e. the arguments of the logs are irrelevant for this notion. Alternatively we call those $r$-fold logarithmic.} For periods coming from closed cycles the highest degree in logarithms of the $l+1$ variables $z_i$ is given by $l-1$, the (complex) dimension of the Calabi-Yau variety. Since in an algebraically realized Calabi-Yau variety, which fixes a polarization, only the primitive part of the horizontal subspace of the middle cohomology can be described by period integrals satisfying Picard-Fuchs equations \cite{MR3965409}, the number of $r$-fold logarithmic solutions corresponds to the dimension $h_{\text{hor prim}}^{(l-1-r,\, r)}$ of the respective piece in that subspace. 
Since the Feynman integral, i.e. the parametrization of the underlying Calabi-Yau variety, is completely symmetric in the $z_i$ variables, solutions always come in complete orbits under permutations of the $z_i$.\footnote{Similarly, the differential equations satisfied by the Feynman integrals always come in complete orbits.} On these grounds it is already possible to propose a generalization of the $l=2,3,4$ results that we explicitly calculated. The number of periods over closed cycles 
is given by
%
\noindent
\begin{table}[htp]
{\centering{
\begin{minipage}[h]{0.55\textwidth}
{\footnotesize{\[
\begin{array}{c}
 \text{\#hol \phantom{hiii} \#log}^1 \\
 \text{\#hol \phantom{hiii}  }\text{\#log}^1\text{  \phantom{hiii} }\text{\#log}^2 \\
 \text{\#hol \phantom{hiii}  }\text{\#log}^1\text{ \phantom{hiii}  }\text{\#log}^2\text{  \phantom{hiii} }\text{\#log}^3 \\
 \text{\#hol \phantom{hiii}  }\text{\#log}^1\text{  \phantom{hiii} }\text{\#log}^2\text{  \phantom{hiii} }\text{\#log}^3\text{  \phantom{hiii} }\text{\#log}^4 \\
 \text{\#hol \phantom{hiii}  }\text{\#log}^1\text{  \phantom{hiii} }\text{\#log}^2\text{ \phantom{hiii}  }\text{\#log}^3\text{ \phantom{hiii}  }\text{\#log}^4\text{  \phantom{hiii} }\text{\#log}^5 \\[1.5ex]
\end{array}
\]}}
\end{minipage}%
\begin{minipage}{0.125\textwidth}
    \begin{tikzpicture}
	        \draw [line width=1.5pt][->] (0,0)--(1.5,0);
    \end{tikzpicture}
\end{minipage}%
\begin{minipage}{0.3\textwidth}
{\footnotesize{\[
\begin{array}{c}
 \text{1\phantom{hiii}1} \\
 \text{1 \phantom{hiii}  4  \phantom{hiii} 1} \\
 \text{1 \phantom{hiii}  5 \phantom{hiii}  5  \phantom{hiii} 1} \\
 \text{1 \phantom{hiii}  6  \phantom{hiii} 15  \phantom{hiii} 6  \phantom{hiii} 1} \\
 \text{1 \phantom{hiii}  7  \phantom{hiii} 21 \phantom{hiii}  21 \phantom{hiii}  7  \phantom{hiii} 1} \\
\end{array}
\]}}
\end{minipage}}}
\caption{The number of logarithmic solution for the non equal mass case for $l=2,\ldots, 6$.}
\label{tablelogsolution}
\end{table}%

\vspace{2ex}

\noindent
Here the entries in the $n$th row correspond to the number $d_{r}^{(l)}$ of basis elements at loop order $l=n+1$ that have the indicated degree $r$ in logarithms, coinciding with the length of an orbit under permutations of the $z_i$. The left side is Pascal's triangle and the right side is fixed by the invariance of the primitive part of the horizontal Hodge numbers\footnote{At that point we want to mention that Fernando Rodriguez Villegas told us that he actually found a motive from which one can directly calculate the horizontal Hodge numbers agreeing with our primitive horizontal Hodge numbers \eqref{primcohom}. We thank him for telling us this fact. Furthermore, we think that with some effort it is also possible to compute the full horizontal Hodge numbers from the complete intersection model described in section \ref{sec:l-loopbanadef}.} of our variety under complex conjugation. 

We shall further illucidate the combinatorial pattern that allows for a complete determation of a Frobenius basis and gives rise to the above numbers. 
After suitable normalization the holomorphic power series starts with unity 
\begin{equation}
	\hat{\varpi}_0(z_i)	=	1	+	\mathcal O(z_i^2)~.
\label{hol}
\end{equation}
For $l>2$ there are $l+1$ single logarithmic solutions of the form
\begin{equation}
	\hat\varpi_1^k(z_i)	=	\log(z_k)	+	\mathcal O(z_i)~.
\label{log}
\end{equation}

Solutions of higher logarithmic degree $r \leq\lceil\frac l2\rceil-1$ are of the form \begin{equation}
	\hat\varpi_r^k(z_i)	=	\prod_{i=1}^r\log(z_{j_i^{(k)}})	+	\mathcal O(z_i)	\qquad \text{for some} \ \ \ \{j_1^{(k)},\hdots, j_r^{(k)} \}\in T^{(l+1)}_r ~,
\label{loglow}
\end{equation}
where $k=1,\hdots,\binom {l+1}r $ now labels the elements of $T^{(l+1)}_r$, which we recall is the set of all subsets of $\{1,\hdots,l+1\}$ of length $r$. %

Further logarithmic solutions for $r > \lceil\frac l2\rceil-1$ are obtained as follows. Label the $\binom{l+1}{l-r-1}$ subsets of $\{ 1, ..., l+1 \}$ having exactly $l-r-1$ elements by $k$. For each $k$, i.e. a choice $N^{(k)} = \{ n^{(k)}_{1}, \dots, n^{(k)}_{l-r-1} \} \subseteq \{ 1, ..., l+1 \}$, the solution $\hat\varpi_r^k$ only involves $r$-fold logarithms in the remaining $r+2$ variables and we have
\begin{equation}
	\hat\varpi_r^k(z_i)	=	\sum_{\{ j_1,\hdots,j_r \} \in \{1,\dots, l+1 \}\backslash N^{(k)} } \, \prod_{i=1}^r\log(z_{j_i})		+	\mathcal O(z_i)~.
\label{loghigh}
\end{equation}
As a consequence of these formulae, the total number of solutions (that correspond to integrals over closed cycles) is given by
\begin{equation}
\sum_{r=0}^{l-1} d^{(l)}_r = 2^{l+1} - \binom{l+2}{\lfloor\frac{l+2}{2}\rfloor}~.
\label{numbersolutions}
\end{equation}

The additional (special) solution can be chosen to start as
\begin{equation}
	\hat\varpi_l(z_i)		=	\prod_{i=1}^{l+1}\log(z_i)\sum_{i=1}^{l+1}\frac1{\log(z_i)}	+	\mathcal O(z_i)~.
\label{logspe}
\end{equation}

\paragraph{A generating function.} As in the equal mass case we can define a generating function for a set of solutions by shifting $k_i \rightarrow k_i + \epsilon_i$ in the series \eqref{maxcutmumnonequal}. Derivatives with respect to the formal parameters $\epsilon_i$ then yield the higher logarithmic solutions. One has to take care that if the degree of the logarithms is larger than one, appropriate linear combinations of various derivatives have to be taken to get a correct solution. That is, these linear combinations should have the same combinatorial structure as the logarithmic solutions in \eqref{loglow}, \eqref{loghigh} and \eqref{logspe}.

\paragraph{Loop reduction.}An interesting feature of the solutions presented here is the following reduction property. Starting from those solutions for the $l$-loop banana integral which have no contribution from $\log(z_{l+1})$, those solutions of the $(l-1)$-loop integral which correspond to closed cycles are obtained from the former by setting $z_{l+1}$ to zero. In this limit some $l$-loop solutions vanish and the number of non-zero logarithmic solutions of the $(l-1)$-loop geometry thus obtained nicely matches the number one expects according to the structure \eqref{hol}-\eqref{loghigh}.

\paragraph{A four-loop example.} To illustrate the structure of the solutions we consider the four-loop case, which is the lowest loop order that, to our knowledge, has not been treated analytically in the non-equal mass case in the literature. Solutions start with
\begin{gleichung}
	\hat\varpi_0(z_i)	&=	1 + 2 \left(z_1 z_2+z_1 z_3+z_2 z_3+z_1 z_4+z_2 z_4+z_3 z_4+z_1 z_5+z_2 z_5	\right.	\\
					&	\qquad \left.+z_3 z_5+z_4 z_5\right) + \mathcal O(z_i)^3	\\
	\hat\varpi_1^1(z_i)	&=	\log \left(z_1\right)-z_1+z_2+z_3+z_4+z_5	+ \mathcal O(z_i)^2	\\
	\hat\varpi_2^1(z_i)	&=	\log \left(z_2\right) \log \left(z_3\right)+\log \left(z_2\right) \log \left(z_4\right)+\log \left(z_3\right) \log \left(z_4\right)+\log \left(z_2\right) \log \left(z_5\right)	\\
	&\qquad+\log
   \left(z_3\right) \log \left(z_5\right)+\log \left(z_4\right) \log \left(z_5\right) 	+	\mathcal O(z_i)\\
	\hat\varpi_3(z_i)	&=	\log \left(z_1\right) \log \left(z_2\right) \log \left(z_3\right)+\log \left(z_1\right) \log \left(z_2\right) \log \left(z_4\right)+\log \left(z_1\right) \log \left(z_3\right) \log
   \left(z_4\right)	\\
   &\qquad +\log \left(z_2\right) \log \left(z_3\right) \log \left(z_4\right)+\log \left(z_1\right) \log \left(z_2\right) \log \left(z_5\right)+\log \left(z_1\right) \log
   \left(z_3\right) \log \left(z_5\right)	\\
   &\qquad+\log \left(z_2\right) \log \left(z_3\right) \log \left(z_5\right)+\log \left(z_1\right) \log \left(z_4\right) \log \left(z_5\right)+\log
   \left(z_2\right) \log \left(z_4\right) \log \left(z_5\right)	\\
   &\qquad+\log \left(z_3\right) \log \left(z_4\right) \log \left(z_5\right)	+ \mathcal O(z_i)	\\
	\hat\varpi_4(z_i)	&=	\log \left(z_1\right) \log \left(z_2\right) \log \left(z_3\right) \log \left(z_4\right)+\log \left(z_1\right) \log \left(z_2\right) \log \left(z_3\right) \log \left(z_5\right)	+ \mathcal O(z_i)\\
	&\qquad+\log \left(z_1\right) \log \left(z_2\right) \log \left(z_4\right) \log \left(z_5\right)+\log \left(z_1\right) \log \left(z_3\right) \log \left(z_4\right) \log \left(z_5\right) \\
	&\qquad +\log
   \left(z_2\right) \log \left(z_3\right) \log \left(z_4\right) \log \left(z_5\right)	+ \mathcal O(z_i)~,
\label{solutions4loop}
\end{gleichung}
where the other single- and double-logarithmic solutions are obtained by replacing $z_1\leftrightarrow z_i$ by $i=2,\hdots,5$.

\paragraph{Analytic continuation by completing the differential ideal.} To extend the solutions $\hat\varpi_n(z_i)$ for $n=0,\hdots,l$ to other domains of the $z_i$-parameter space analytic continuation is needed. To this end it is necessary to have a complete set 
 of differential equations. By these we mean a set of differential equations such that the number of corresponding solutions is equal to the number of solutions given in \eqref{numbersolutions} plus the additional special solution. Notice that in general only the total number of solutions stays the same upon analytic continuation to other points. The precise logarithmic structure of the solutions changes, in particular for analytic continuation to non-singular points of the differential equations.
With this (or these) additional differential equation(s) one can transform the local solutions of the MUM point to domains beyond the original domain of convergence by matching local Frobenius bases on overlapping regions.\footnote{Of course, if one had again sufficient information about the analytic structure of a local Frobenius basis in the new region, one could construct a basis despite only knowing an incomplete set of differential equations, as was the case for the MUM point.}

 As we have explicitly seen in the two- and three-loop case\footnote{For the two- and three-loop case these (inhomogeneous) differential equations are listed in \cite{Klemm:2019dbm} and in the auxiliary \texttt{mathematica}-file accompanying this paper, see \url{http://www.th.physik.uni-bonn.de/Groups/Klemm/data.php}.} some differential equations get even extended with inhomogeneities if the whole Feynman integral $\hat{\mathcal F}_{\sigma_l}$ should satisfy them instead of just the maximal cut $\hat{\mathcal F}_{T^l}$. For the four-loop case we have found a second order operator with coefficients being polynomials of multidegree three in the five variables $z_i$, leading to an inhomogeneous differential equation given in appendix~\ref{sec:appop}. It is hard to give a general formula for the additional and perhaps inhomogeneous differential equation(s). In fact, it is not even clear whether one has to extend the operators $\mathcal D_k$ just by a single differential equation with or without an inhomogeneity. The general strategy to get a complete system of (inhomogeneous) differential equations is simply that one searches for new ones until only the expected number of solutions is determined by these operators. We suggest to search for second order operators by systematically increasing the multidegree of the coefficient polynomials multiplying the derivatives. At some point these are expected to yield a complete set of differential equations, as we checked for $l\leq 4$. If not, one has to go to higher degree equations in the ansatz.

\subsection{Linear combination for the non-equal mass Feynman integral}

Next we fix the linear combination of the previously constructed solutions that gives the full non-equal mass Feynman integral
\begin{gleichung}
	\hat{\mathcal F}_{\sigma_l}	=	\sum_{r=0}^{l} \sum_{s=1}^{d^{(l)}_r} \lambda^{(l)}_{r,s} \, \hat\varpi_r^s~.	
\label{lincomnonequal}
\end{gleichung}
As in the equal mass case we numerically compute $\hat{\mathcal F}_{\sigma_l}$ to fix the coefficients $\lambda_{r,s}^{(l)}$. These  again turn out to be appropriate combinations of zeta values, closely related to the equal mass ones $\lambda_r^{(l)}$. For $l=2,3,4$ and with respect to the basis of solutions given by \eqref{hol}, \eqref{log}, \eqref{loglow},\eqref{loghigh} and \eqref{logspe} we find the explicit values shown in Table \ref{tablambdanonequal}.
\begin{table}[htp]
\begin{center}\begin{tabular}{c?ccccc}
	$l $&$ \lambda _{0,s}{}^{(l)} $&$ \lambda _{1,s}{}^{(l)} $&$ \lambda _{2,s}{}^{(l)} $&$ \lambda_{3,s}{}^{(l)} $&$ \lambda _{4,s}{}^{(l)}$ \\[1ex]\toprule[1.5pt]
 	$2 $&$ 18 \text{$\zeta $(2)} $&$ 2 \pi  i $&$ 1 $&$ \text{} $&$ \text{}$ \\[1ex]
	$3 $&$ -16 \text{$\zeta $(3)+24$\pi \zeta $(2)}i $&$ 
\begin{array}{c}
 -18 \text{$\zeta $(2)} \\
 -18 \text{$\zeta $(2)} \\
 -18 \text{$\zeta $(2)} \\
 -18 \text{$\zeta $(2)} \\
\end{array}
		 $&$ -2 \pi  i $&$ 1 $&$ \text{}$ \\[8ex]
  	$4 $&$ -450 \text{$\zeta $(4)-80$\pi \zeta $(3)}i $&$ 
\begin{array}{c}
 16 \text{$\zeta $(3)}$-24$\text{$\pi \zeta $(2)}i \\
 16 \text{$\zeta $(3)}$-24$\text{$\pi \zeta $(2)}i \\
 16 \text{$\zeta $(3)}$-24$\text{$\pi \zeta $(2)}i \\
 16 \text{$\zeta $(3)}$-24$\text{$\pi \zeta $(2)}i \\
 16 \text{$\zeta $(3)}$-24$\text{$\pi \zeta $(2)}i \\
\end{array}
 		$&$ 
\begin{array}{c}
 6 $\text{$\zeta $(2)}$ \\
 6 $\text{$\zeta $(2)}$ \\
 6 $\text{$\zeta $(2)}$ \\
 6 $\text{$\zeta $(2)}$ \\
 6 $\text{$\zeta $(2)}$ \\
\end{array}
		 $&$ 2 \pi  i $&$ 1$ \\
\end{tabular}
\end{center}
\caption{Coefficients giving the non-equal mass Feynman integral in the MUM point Frobenius basis of subsection \ref{ssc:non-eq-mass-diff}. }
\label{tablambdanonequal}
\end{table}%
\noindent
We see that with our choice of basis the equal mass values for $\lambda_r^{(l)}$ split symmetrically into the values $\lambda_{r,s}^{(l)}$. In general we claim that the non-equal mass values satisy
\begin{equation}\label{lambdanonequalclaim}
	\lambda_{r,s}^{(l)}	= \begin{cases}   \lambda_r^{(l)} \cdot	\binom{l+1}{r}^{-1}& \text{for } r\leq\lceil\frac l2\rceil-1 \text{ and } s=1,\hdots,\binom{l+1}{r} \\  \lambda_r^{(l)}\cdot 
	\left(\binom{l+1}{l-r-1}\binom{r+2}{r}\right)^{-1} & \text{for } r> \lceil\frac l2\rceil-1 \text{ and }s=1,\hdots,\binom{l+1}{l-r-1} \\
	1 & \text{for } r=l\qquad \ \ \, \text{ and } s=1~.
	 \end{cases} 
\end{equation}
We remark that the factor $\binom{r+2}r$ results from the $\binom{r+2}r$ terms in the sum of \eqref{loghigh}.

\subsection{Remarks about master integrals for generalized banana Feynman integrals} 
\label{masterintegrals}

Finally, we want to connect our results to certain master integrals for a (different kind of) family of banana type Feynman integrals. The latter family not only includes the Feynman integral $\mathcal F_{\sigma_l}$ with \textit{fixed} loop order $l$, but also all integrals obtained by raising the propagators in the denominator to some non-negative powers $\nu_i$ and/or including polynomials in dot products of (external or loop) momenta in the numerator of the momentum space integrand. This constitutes a generalization of the kind of integrals considered so far in this work. By master integrals we then mean a finite subset of these integrals, i.e., a set of choices for the powers of propagators and powers of dot products, such that all other integrals in the family are obtained by linearly combining the master integrals,  where the coefficients in general are rational functions\footnote{The precise definition of master integrals is immaterial at this points, as the only claims made will concern a set of integrals that, as we believe, should be amongst the master integrals in any reasonable defintion.} in the kinematic parameters (external momenta and masses). 

For the two- and three-loop equal mass banana integrals sets of master integrals are known (see for instance \cite{Laporta:2004rb} and \cite{Primo:2017ipr}). In the non-equal mass case less is known. To the best of our knowledge only in the two-loop case all master integrals and their relations to the other integrals where found \cite{Caffo:1998du}. At least results for the number of higher loop  master integrals are available in \cite{Kalmykov:2016lxx,Bitoun:2017nre}.

Given our solution for $\mathcal F_{\sigma_l}$ we can at least construct a (sub-)set of master integrals which is possibly neither complete nor linearly independent\footnote{Here again linear dependence refers to coefficients being rational functions in the kinematic parameters.}, namely those integrals with trivial dot products in the numerator (of the standard momentum space represenation) but positive powers of the propagators in the denominator. For this note that raising the $k{\text{th}}$ propagator to the power $\nu_k$ means taking the $(\nu_k-1){\text{th}}$ derivative of the original Feynman integral $\mathcal F_{\sigma_l}$ (where all $\nu_i=1$) with respect to the mass parameter $\xi_k^2$, i.e.
\begin{gleichung}
	\mathcal F_{\sigma_l}(t,\xi_i;\nu_1,\hdots,\nu_{l+1})	=	\prod_{k=1}^{l+1}\partial_{\xi_k^2}^{\nu_k-1} \mathcal F_{\sigma_l}(t,\xi_i;1,\hdots,1)	=	\prod_{k=1}^{l+1}\partial_{\xi_k^2}^{\nu_k-1}\mathcal F_{\sigma_l}(t,\xi_i)~.
\label{relationmaster}
\end{gleichung}
Since we have fixed the linear combination of the MUM-point Frobenius basis yielding the original Feynman integral $\mathcal F_{\sigma_l}$, we also have the correct linear combination for the mentioned (master) integrals $\mathcal F_{\sigma_l}(t,\xi_i;\nu_1,\hdots,\nu_{l+1})$ by the relation \eqref{relationmaster}. As a consequence, expansions in the masses $\xi_i^2$ and the momentum $t$ are readily available.

For the master integrals in the equal mass case we expect less independent functions. Of course, one can first construct the non-equal mass banana master integrals and at the end restrict to the equal mass case by setting all masses to unity. However, it is actually much simpler to consider only the derivatives
\begin{equation}
	\frac1{r!}\left[ \prod_{k=1}^{r}(\theta_t+k)\right] \mathcal F_{\sigma_l}(t,1)~,
\end{equation} 
for at least $r\leq l-1$. These derivatives correspond to the integrals
\begin{gleichung}
	\int_{\sigma_l}	\frac{\mathcal U^r}{\mathcal F^{r+1}}\left(\sum_{k=1}^{l+1}x_i\right)^r\mu_l~,
\end{gleichung}
where 
\begin{equation}
 \mathcal U=\prod_{k=1}^{l+1}x_k\sum_{k=1}^{l+1}\tfrac1{x_k} \qquad \text{and} \qquad \mathcal F=\mathcal U \sum_{k=1}^{l+1}x_k\xi_k^2-t\prod_{k=1}^{l+1}x_k
\end{equation} are the two Symanzik polynomials for the banana graphs. We expect that these integrals form part\footnote{At least one has to extend this set of master integrals by the constant function corresponding to the tadpole integral, which arises as a subtopology of the banana graph.} of the basis for the equal mass $l$-loop family of banana integrals.

\section{Conclusion and outlook}
\label{sec:conclusion}
In this paper  we used techniques  developed in the interface of algebraic 
geometry and algebraic number theory with mathematical physics mostly in 
the context of string theory to describe the complete analytic structure of the
$l$-loop banana integrals. While the low energy region was relatively  well under 
control using the Bessel function integrals most of the results near  the region 
$t \ge (\sum_{i=1}^{l+1} \xi_i)^2$ and into the high energy regime are new and 
related in a simple and beautiful way to the realization of the Feynman 
amplitudes as periods of very symmetric complete intersection  Calabi-Yau 
$(l-1)$-folds and their extension.  

In finding the most suitable geometry it turned out that the motivic perspective 
is as important as the geometrical intuition. So one lesson to keep in mind 
is to try first to  find the motive based on  any available exact period that one can get 
hands on, in the simplest possible or most suitable geometric realizations for the 
question at hand. For this task data bases for motives of Calabi-Yau-- and 
Fano motives are very useful. 

Due to the $\widehat \Gamma$-class evaluation the occurrence of products of zeta values in the large energy limit, whose 
highest degree of transcendentality is $l$, is by now explained   
for the banana graphs. However, while the numerical evidence is overwhelming  
our application of the $\widehat \Gamma$-class to the Fano variety (somewhat different as to the 
Calabi-Yau  to get the imaginary part)  is still conjectural and poses  an interesting  
challenge to prove it in mathematical rigor. Nevertheless, the convincing evidence   
might shed light  on the general observation that Feynman integrals evaluate at special 
points to interesting (conjecturally transcendental) numbers, such as multiple zeta values (MZV) or 
critical values of modular L-functions. For the latter fact we have found an interesting new 
example at the attractor points of the maximal cut integral. At this point the Galois representation 
splits relative to the generic Calabi-Yau family into two simple factors.  

It might be interesting  to relate this to more systematic 
studies  of the  degree of transcendentality and the motivic Galois group. For example   for massless 
$\phi^4$-theory, a specific kind of Feynman periods is studied in~\cite{Schnetz:2016fhy}, 
originating from primitive vacuum graphs --- more precisely, primitive logarithmically divergent 
graphs (with external legs) may be regularised, and the residue in the regulator is 
in~\cite{Schnetz:2016fhy} called the period of the graph (with amputated legs). This sort of 
period gives a renormalization scheme independent contribution to the $\beta$-function~\cite{Itzykson:1980rh}. 
These massless $\phi^4$-periods have also served as a data mine for exploring the number theory 
content of perturbative QFT. Notably, the family of zig-zag graphs gives the only family of periods 
known to all loop orders, proven to be rational multiples of odd zeta values \cite{Brown:2012ia}. 
However, the number content of QFTs exceeds the span of multiple zeta values 
\cite{Brown:2013wn,Brown:2010bw}. There is a conjectured Galois coaction on the 
periods \cite{Panzer:2016snt,Schnetz:2017bko}, compatible with known $\phi^4$-periods 
and also with the number content \cite{Laporta:2017okg} of the electron anomalous 
magnetic moment in QED. 

The most practical consequence of the present work is that the GKZ methods and the understanding of the analytic structure of the amplitudes allow to calculate it extremely fast and to very high 
numerical  precision or as simple series expansion. A first analysis in subsection~\ref{masterintegrals} indicates that this is also true for the master integrals that are necessary to dwell in actual experimental data analysis. Furthermore, it is reasonable to have similar expectations for the period integrals of higher orders in the dimensional regularization parameter $\varepsilon$\footnote{We are tempted to speculate on the existence of differential operators in the moduli relating the $\varepsilon^0$-order integals (i.e. $d=2$) computed in this work and the higher terms in $\varepsilon$, similar to the calculations in \cite{Huang:2014nwa,Fischbach:2018yiu}.}. It would be interesting to collaborate with complementary expertise 
to  provide actual program tools for this task.

The very richly nested but, in each occasion, simple fibration structure going back all the way to the elliptic fibrations  
of the K3 surfaces in the geometry describing the Feynman amplitude, as mentioned in subsection \ref{sec:gammaclass},  guarantees 
that there are all kinds of overlapping limits in which simpler functions, as for example the elliptic dilogarithm, 
are bound to occur~\cite{Bloch:2013tra,Broedel:2019kmn}.  In the context of large base parameter of  K3 fibered Calabi-Yau 
three-folds it is known by heterotic Type II duality that exact modular functions obtained by the Borcherds 
lift do describe the integral instanton expansion completely~\cite{Klemm:1995tj,Marino:1998pg,Klemm:2005pd,Grimm:2007tm}.
This suggests that there will be a further  confluence of techniques to understand more generally the automorphic and 
integral structures of the Feynman amplitudes.

\acknowledgments
It is a pleasure to thank Don Zagier for finding the generating functions for $\lambda$-values, commenting on the 
double Borel sum of the symmetric product of Bessel functions and collaboration on work related to periods and 
quasiperiods of modular forms. We further want to thank the members of the {\sl  { International Group de Travail on differential equations}} (in Paris), initiated by Vasily 
Golyshev to create a lively research environment during the pandemic,  for most important remarks. In particular, 
Matt Kerr for crucial discussions about the geometry of the three-loop graph,  
to Fernando Rodriguez Villegas for providing an alternative motive for the periods over the primitive cohomology and 
Danylo Radchenko for sending us \texttt{Sage} code to calculate the corresponding Ehrhart-polynomials. To 
Claude Sabbah, Matt Kerr and Spencer Bloch  for pointing out state of the art knowledge about the special values 
at $t=1$ using L-functions and Kloosterman sums. Special thanks to Claude Duhr for providing the physical 
motivations  and  a most valuable first hand perspective from the practitioner. We also would like to thank Hiroshi Iritani for a private communication confirming the validity of formula \eqref{K-theorychain}. Moreover, we want to thank Victor Batyrev, Yang-Hui He, Philip Candelas, 
Xenia de la Ossa, Mohamed Elmi, Amir Kashani-Poor, Duco van Straten and Hans Jockers for further help and discussions. Kilian B\"onisch thanks the German Academic Scholarship Foundation for financial support. We also thank the Bonn-Cologne Graduate School of Physics and Astronomy for financial support.



\newpage
\appendix

\section{Derivation of the Bessel function representation}
\label{appbessel}

In this appendix we rewrite the Feynman integral \eqref{bananageneral} in terms of an integral over Bessel functions, closely follow the discussion in \cite{Vanhove:2014wqa}.

Starting from \eqref{bananageneral} one expands a factor in the denominator in terms of a geometric series,
\begin{align}
	{\cal F}_{\sigma_l}	
		&=	-\sum_{k=0}^\infty t^k\int_{\sigma_l} \left(	\frac{1}{\left(\sum_i \xi_i^2 x_i\right) \left(\sum_i x_i^{-1}\right)}	\right)^{k+1}~ \frac{\mu_l}{\prod_i x_i }	~,
\label{2}
\end{align}
converging if $t 	<	\left(\sum_i\xi_i^2x_i\right)\left(\sum_i\frac{1}{x_i}\right)$. Since $x_i \geq0$ we require
\begin{gleichung}
	t	<	\left(\sum_{i=1}^{l+1}\xi_i\right)^2
\end{gleichung}
 for equation \eqref{2} to hold. Furthermore, we can use the identity
\begin{gleichung}
	\left(\frac1a\right)^{k+1}	=	\frac1{k!}\int_0^\infty\mathrm e^{-ax}x^k~\mathrm dx~,
\end{gleichung}
which is valid for $\text{Re}(a)>0$ and $k>-1$, to rewrite the denominator in \eqref{2} introducing two new integrations
\begin{gleichung}
	{\cal F}_{\sigma_l}	=	-\sum_{k=0}^\infty \frac{t^k}{(k!)^2}\int_{\sigma_l}\int_0^\infty\int_0^\infty\mathrm e^{-u\sum_i\xi_i^2x_i-v\sum_i\frac1{x_i}}~\frac{\mathrm du  \mathrm dv}{u^{-k}v^{-k}}\frac{\mu_l}{\prod_i x_i }~.	
\end{gleichung}
The projective integral over $\sigma_l$ can be performed using the identity
\begin{gleichung}
	\int_0^\infty \mathrm e^{-u m^2x-\frac vx}~\frac{\mathrm dx}x	=	2K_0(2m\sqrt{uv})
\end{gleichung}
involving the Bessel function of the second kind $K_0$. We obtain
\begin{gleichung}
	{\cal F}_{\sigma_l}	=	-2^l\sum_{k=0}^\infty\frac{t^k}{(k!)^2}\int_0^\infty\int_0^\infty\prod_{i=1}^l K_0\left(2\xi_i \sqrt{uv}\right) \mathrm e^{-u\xi_{l+1}^2-v}~ \frac{\mathrm du\mathrm dv}{u^{-k}v^{-k}}~.
\end{gleichung}
Introducing new variables $y = v$ and $z = 2\sqrt{uv}$ with $\dd u\dd v=\frac{z}{2y}~\dd y\dd z$ and integrating subsequently over $y$ we find
\begin{gleichung}
	{\cal F}_{\sigma_l}	=	-2^{l+1}\sum_{k=0}^\infty\frac{t^k}{(k!)^2}\int_0^\infty\prod_{i=1}^{l+1} K_0(2\xi_i) \left(\frac{z}{2}\right)^{2k+1}~\dd z~.
\end{gleichung}
The Bessel function of the first kind $I_0$ has a series representation given by
\begin{gleichung}
	I_0(x)	=	\sum_{k=0}^\infty \left(\frac x2\right)^{2k}\frac1{(k!)^2}~,	
\end{gleichung}
which simplifies ${\cal F}_{\sigma_l}$ to the final expression
\begin{gleichung}
	{\cal F}_{\sigma_l}	=	2^l\int_0^\infty z \, I_0(\sqrt t z) \prod_{i=1}^{l+1}K_0(\xi_i z)~ \dd z~.
\end{gleichung}	

\section{Inhomogeneous differential equation for the four-loop case }
\label{sec:appop}

\noindent In this appendix we give an inhomogeneous differential equation
\begin{equation}\label{eq:Fs4inhom}
\mathcal D \hat{\mathcal{F}}_{\sigma_4} = S
\end{equation}
  satisfied by the four-loop banana Feynman integral $\hat{\mathcal{F}}_{\sigma_4}$ (which includes the extra factor \eqref{eq:a0} in the numerator) in the case of generic masses. Operators $\mathcal D_k$ leading to homogeneous differential equations for $\hat{\mathcal{F}}_{\sigma_4}$ have already been given in \eqref{homops}. 
Indeed, here we only present the leading contribution in $z_i$ to $\mathcal D$, which reads
\begin{gleichung} 
	\mathcal D	&=	-63 \, \theta _2^2-416 \, \theta _1 \theta _3-13 \, \theta _2 \theta _3+206 \, \theta _3^2-180 \, \theta _1 \theta _4+102 \, \theta _2 \theta _4+507 \, \theta _3 \theta _4+180\, \theta _4^2	\\
   	&\qquad +596 \, \theta _1 \theta_5-89 \, \theta _2 \theta _5-78 \, \theta _3 \theta _5-429 \, \theta _4 \theta _5-323 \, \theta _5^2	 + \mathcal O(z_i)~. \label{inhomopfourloop}
\end{gleichung}
The complete expression of this second-order operator $\mathcal D$ can be found in a supplementary \texttt{mathematica}-file on our web page\footnote{\url{http://www.th.physik.uni-bonn.de/Groups/Klemm/data.php}}. Furthermore, the inhomogeneity to $\mathcal D$ is given by
\begin{gleichung}
 S&=\	\left(-42 z_1+168 z_2-101 z_3+282 z_4-139 z_5\right) \log \left(z_1\right) \log \left(z_2\right)\\
   	&\qquad+\left(-416-556 z_1+283 z_2+105 z_3-15 z_4-128 z_5\right) \log \left(z_1\right) \log\left(z_3\right)\\
   	&\qquad+\left(-180-180 z_1-345 z_2-195 z_3+540 z_5\right) \log \left(z_1\right) \log \left(z_4\right)\\
   	&\qquad+\left(596+778 z_1-106 z_2+191 z_3-267 z_4-273 z_5\right) \log \left(z_1\right) \log \left(z_5\right)\\
   	&\qquad+\left(-13+533 z_1+203 z_2-21 z_3-15 z_4-128 z_5\right) \log \left(z_2\right) \log \left(z_3\right)\\
   	&\qquad+\left(102+123 z_1+168 z_2-195 z_3-6 z_5\right) \log \left(z_2\right) \log \left(z_4\right)\\
   	&\qquad+\left(-89-614 z_1-539 z_2+317 z_3-267 z_4+273 z_5\right) \log \left(z_2\right) \log \left(z_5\right)\\
   	&\qquad+\left(507+122 z_1-477 z_2+407 z_3-252 z_4-139 z_5\right) \log\left(z_3\right) \log \left(z_4\right)\\
   	&\qquad+\left(-78-99 z_1-9 z_2-491 z_3+282 z_4+395 z_5\right) \log \left(z_3\right) \log \left(z_5\right)\\
   	&\qquad+\left(-429-65 z_1+654 z_2-17 z_3+252 z_4-395z_5\right) \log \left(z_4\right) \log \left(z_5\right)~. \hspace{5cm}
\label{inhomfourloop}
\end{gleichung}

\section{\texttt{Pari/GP} script for equal mass amplitude}
\label{sec:pariprogram}

The \texttt{Pari/GP} script \texttt{BananaAmplitude.gp}\footnote{The script can be downloaded on our web page \url{http://www.th.physik.uni-bonn.de/Groups/Klemm/data.php}.} allows to compute the equal mass amplitude for any given loop order and to any given precision. To explain how one can use the program we want to numerically confirm the special value of the four-loop amplitude at $s=1$.

After reading the script into \texttt{GP} one has to specify the number of significant digits for internal computations and a variable $N$ that controls to which order the solutions of the differential equations are expanded:
\begin{verbatim}
? \p 60
   realprecision = 77 significant digits (60 digits displayed)
? N=60;
\end{verbatim}
These variables have to be chosen by hand to match the desired accuracy. A linear change in the variable that limits the accuracy results in a linear change of the accuracy in decimal digits\footnote{One possibility for checking the accuracy is to evaluate the amplitude for some $s>1/(l+1)^2$. In this regime the imaginary part has to vanish and this gives a good method to control the accuracy.}. To generate the $l$-loop amplitude one has to call the associated function and give the number of loops.
\begin{verbatim}
? f=AMPLITUDE(4);
\end{verbatim}
In this step, the solutions of the differential equation are expanded around all singular points and the amplitude is analytically continued from $s=0$ to the complete positive real axis. The amplitude $f(s)$ can now be calculated for any positive $s \neq 1/(l+1)^2$. E.g.\ to check the special value at $s=1$ we write:
\begin{Verbatim}[fontsize=\small]
? f(1)-8*Pi^2*lfun(lfunmf(mfinit([6,4],0),mffrometaquo([1,2;2,2;3,2;6,2])),2)
%11 = -2.18604340588457524292517561761512202353 E-52 
       + 4.9173920300900275825221994294591114746 E-53*I
\end{Verbatim}

\newpage
\bibliographystyle{JHEP}
\bibliography{References}

\providecommand{\href}[2]{#2}\begingroup\raggedright\begin{thebibliography}{10}

\bibitem{Chetyrkin:1981qh}
K.~Chetyrkin and F.~Tkachov, \emph{{Integration by Parts: The Algorithm to
  Calculate beta Functions in 4 Loops}},
  \href{https://doi.org/10.1016/0550-3213(81)90199-1}{\emph{Nucl. Phys. B}
  {\bfseries 192} (1981) 159}.

\bibitem{Tkachov:1981wb}
F.~Tkachov, \emph{{A Theorem on Analytical Calculability of Four Loop
  Renormalization Group Functions}},
  \href{https://doi.org/10.1016/0370-2693(81)90288-4}{\emph{Phys. Lett. B}
  {\bfseries 100} (1981) 65}.

\bibitem{Grozin:2011mt}
A.~Grozin, \emph{{Integration by parts: An Introduction}},
  \href{https://doi.org/10.1142/S0217751X11053687}{\emph{Int. J. Mod. Phys. A}
  {\bfseries 26} (2011) 2807}
  [\href{https://arxiv.org/abs/1104.3993}{{\ttfamily 1104.3993}}].

\bibitem{Zhang:2016kfo}
Y.~Zhang, \emph{{Lecture Notes on Multi-loop Integral Reduction and Applied
  Algebraic Geometry}},  12, 2016,
  \href{https://arxiv.org/abs/1612.02249}{{\ttfamily 1612.02249}}.

\bibitem{Smirnov:2012gma}
V.~A. Smirnov, \emph{{Analytic tools for Feynman integrals}}, vol.~250. 2012,
  \href{https://doi.org/10.1007/978-3-642-34886-0}{10.1007/978-3-642-34886-0}.

\bibitem{Bauberger:1994nk}
S.~Bauberger, M.~B{\"o}hm, G.~Weiglein, F.~A. Berends and M.~Buza,
  \emph{{Calculation of two-loop self-energies in the electroweak Standard
  Model}}, \href{https://doi.org/10.1016/0920-5632(94)90665-3}{\emph{Nucl.
  Phys. B Proc. Suppl.} {\bfseries 37B} (1994) 95}
  [\href{https://arxiv.org/abs/hep-ph/9406404}{{\ttfamily hep-ph/9406404}}].

\bibitem{Bonciani:2019jyb}
R.~Bonciani, V.~Del~Duca, H.~Frellesvig, J.~Henn, M.~Hidding, L.~Maestri
  et~al., \emph{{Evaluating a family of two-loop non-planar master integrals
  for Higgs + jet production with full heavy-quark mass dependence}},
  \href{https://doi.org/10.1007/JHEP01(2020)132}{\emph{JHEP} {\bfseries 01}
  (2020) 132} [\href{https://arxiv.org/abs/1907.13156}{{\ttfamily
  1907.13156}}].

\bibitem{Abreu:2019fgk}
S.~Abreu, M.~Becchetti, C.~Duhr and R.~Marzucca, \emph{{Three-loop
  contributions to the $\rho$ parameter and iterated integrals of modular
  forms}}, \href{https://doi.org/10.1007/JHEP02(2020)050}{\emph{JHEP}
  {\bfseries 02} (2020) 050}
  [\href{https://arxiv.org/abs/1912.02747}{{\ttfamily 1912.02747}}].

\bibitem{Laporta:2008sx}
S.~Laporta, \emph{{Analytical expressions of 3 and 4-loop sunrise Feynman
  integrals and 4-dimensional lattice integrals}},
  \href{https://doi.org/10.1142/S0217751X08042869}{\emph{Int. J. Mod. Phys. A}
  {\bfseries 23} (2008) 5007}
  [\href{https://arxiv.org/abs/0803.1007}{{\ttfamily 0803.1007}}].

\bibitem{Adams:2016xah}
L.~Adams, C.~Bogner, A.~Schweitzer and S.~Weinzierl, \emph{{The kite integral
  to all orders in terms of elliptic polylogarithms}},
  \href{https://doi.org/10.1063/1.4969060}{\emph{J. Math. Phys.} {\bfseries 57}
  (2016) 122302} [\href{https://arxiv.org/abs/1607.01571}{{\ttfamily
  1607.01571}}].

\bibitem{Broedel:2019kmn}
J.~Broedel, C.~Duhr, F.~Dulat, R.~Marzucca, B.~Penante and L.~Tancredi,
  \emph{{An analytic solution for the equal-mass banana graph}},
  \href{https://doi.org/10.1007/JHEP09(2019)112}{\emph{JHEP} {\bfseries 09}
  (2019) 112} [\href{https://arxiv.org/abs/1907.03787}{{\ttfamily
  1907.03787}}].

\bibitem{Kalmykov:2016lxx}
M.~Y. Kalmykov and B.~A. Kniehl, \emph{{Counting the number of master integrals
  for sunrise diagrams via the Mellin-Barnes representation}},
  \href{https://doi.org/10.1007/JHEP07(2017)031}{\emph{JHEP} {\bfseries 07}
  (2017) 031} [\href{https://arxiv.org/abs/1612.06637}{{\ttfamily
  1612.06637}}].

\bibitem{Bitoun:2017nre}
T.~Bitoun, C.~Bogner, R.~P. Klausen and E.~Panzer, \emph{{Feynman integral
  relations from parametric annihilators}},
  \href{https://doi.org/10.1007/s11005-018-1114-8}{\emph{Lett. Math. Phys.}
  {\bfseries 109} (2019) 497}
  [\href{https://arxiv.org/abs/1712.09215}{{\ttfamily 1712.09215}}].

\bibitem{Adams:2017ejb}
L.~Adams and S.~Weinzierl, \emph{{Feynman integrals and iterated integrals of
  modular forms}},
  \href{https://doi.org/10.4310/CNTP.2018.v12.n2.a1}{\emph{Commun. Num. Theor.
  Phys.} {\bfseries 12} (2018) 193}
  [\href{https://arxiv.org/abs/1704.08895}{{\ttfamily 1704.08895}}].

\bibitem{Ablinger:2017bjx}
J.~Ablinger, J.~Bl\"umlein, A.~De~Freitas, M.~van Hoeij, E.~Imamoglu, C.~Raab
  et~al., \emph{{Iterated Elliptic and Hypergeometric Integrals for Feynman
  Diagrams}}, \href{https://doi.org/10.1063/1.4986417}{\emph{J. Math. Phys.}
  {\bfseries 59} (2018) 062305}
  [\href{https://arxiv.org/abs/1706.01299}{{\ttfamily 1706.01299}}].

\bibitem{Klemm:2019dbm}
A.~Klemm, C.~Nega and R.~Safari, \emph{{The $l$-loop Banana Amplitude from GKZ
  Systems and relative Calabi-Yau Periods}},
  \href{https://doi.org/10.1007/JHEP04(2020)088}{\emph{JHEP} {\bfseries 04}
  (2020) 088} [\href{https://arxiv.org/abs/1912.06201}{{\ttfamily
  1912.06201}}].

\bibitem{Hosono:1994ax}
S.~Hosono, A.~Klemm, S.~Theisen and S.-T. Yau, \emph{{Mirror symmetry, mirror
  map and applications to complete intersection Calabi-Yau spaces}},
  \href{https://doi.org/10.1016/0550-3213(94)00440-P}{\emph{AMS/IP Stud. Adv.
  Math.} {\bfseries 1} (1996) 545}
  [\href{https://arxiv.org/abs/hep-th/9406055}{{\ttfamily hep-th/9406055}}].

\bibitem{MR3965409}
A.~Klemm, \emph{The {B}-model approach to topological string theory on
  {C}alabi-{Y}au n-folds},  in \emph{B-model {G}romov-{W}itten theory}, Trends
  Math., pp.~79--397, Birkh\"{a}user/Springer, Cham, (2018).

\bibitem{Bogner:2007mn}
C.~Bogner and S.~Weinzierl, \emph{{Periods and Feynman integrals}},
  \href{https://doi.org/10.1063/1.3106041}{\emph{J. Math. Phys.} {\bfseries 50}
  (2009) 042302} [\href{https://arxiv.org/abs/0711.4863}{{\ttfamily
  0711.4863}}].

\bibitem{MR1852188}
M.~Kontsevich and D.~Zagier, \emph{Periods},  in \emph{Mathematics
  unlimited---2001 and beyond}, pp.~771--808, Springer, Berlin, (2001).

\bibitem{MullerStach:2012mp}
{M\"{u}ller-Stach, Stefan and Weinzierl, Stefan and Zayadeh, Raphael},
  \emph{{Picard-Fuchs equations for Feynman integrals}},
  \href{https://doi.org/10.1007/s00220-013-1838-3}{\emph{Commun. Math. Phys.}
  {\bfseries 326} (2014) 237}
  [\href{https://arxiv.org/abs/1212.4389}{{\ttfamily 1212.4389}}].

\bibitem{Vanhove:2014wqa}
P.~Vanhove, \emph{{The physics and the mixed Hodge structure of Feynman
  integrals}}, \href{https://doi.org/10.1090/pspum/088/01455}{\emph{Proc. Symp.
  Pure Math.} {\bfseries 88} (2014) 161}
  [\href{https://arxiv.org/abs/1401.6438}{{\ttfamily 1401.6438}}].

\bibitem{MR2810322}
D.~A. Cox, J.~B. Little and H.~K. Schenck, \emph{Toric varieties}, vol.~124 of
  \emph{Graduate Studies in Mathematics}. American Mathematical Society,
  Providence, RI, 2011,
  \href{https://doi.org/10.1090/gsm/124}{10.1090/gsm/124}.

\bibitem{Vanhove:2018mto}
P.~Vanhove, \emph{{Feynman integrals, toric geometry and mirror symmetry}},  in
  \emph{{KMPB Conference}: {Elliptic Integrals, Elliptic Functions and Modular
  Forms in Quantum Field Theory}}, pp.~415--458, 2019,
  \href{https://doi.org/10.1007/978-3-030-04480-0\_17}{DOI}
  [\href{https://arxiv.org/abs/1807.11466}{{\ttfamily 1807.11466}}].

\bibitem{MR1020882}
{Gel'fand, I. M. and Zelevinsky, A. V. and Kapranov, M. M.}, \emph{Newton
  polyhedra of principal {$A$}-determinants}, {\emph{Dokl. Akad. Nauk SSSR}
  {\bfseries 308} (1989) 20}.

\bibitem{MR1080980}
{Gel'fand, I. M. and Kapranov, M. M. and Zelevinsky, A. V.}, \emph{Generalized
  {E}uler integrals and {$A$}-hypergeometric functions},
  \href{https://doi.org/10.1016/0001-8708(90)90048-R}{\emph{Adv. Math.}
  {\bfseries 84} (1990) 255}.

\bibitem{MR1011353}
{Gel'fand, I. M. and Zelevinsky, A. V. and Kapranov, M. M.},
  \emph{Hypergeometric functions and toric varieties},
  \href{https://doi.org/10.1007/BF01078777}{\emph{Funktsional. Anal. i
  Prilozhen.} {\bfseries 23} (1989) 12}.

\bibitem{NasrollahpPeriodsFeynmanDiagrams2016}
E.~Nasrollahpoursamami, \emph{Periods of {{Feynman Diagrams}} and {{GKZ
  D}}-{{Modules}}},  \href{https://arxiv.org/abs/1605.04970}{{\ttfamily
  1605.04970}}.

\bibitem{Nasrollahpoursamami:2017shc}
E.~Nasrollahpoursamami, \emph{{Periods of Feynman diagrams}}, Ph.D. thesis,
  Caltech, Pasadena (main), 2017.
\newblock 10.7907/Z9GX48MR.

\bibitem{Feng:2019bdx}
T.-F. Feng, C.-H. Chang, J.-B. Chen and H.-B. Zhang, \emph{{GKZ-hypergeometric
  systems for Feynman integrals}},
  \href{https://doi.org/10.1016/j.nuclphysb.2020.114952}{\emph{Nucl. Phys. B}
  {\bfseries 953} (2020) 114952}
  [\href{https://arxiv.org/abs/1912.01726}{{\ttfamily 1912.01726}}].

\bibitem{Klausen:2019hrg}
R.~P. Klausen, \emph{{Hypergeometric Series Representations of Feynman
  Integrals by GKZ Hypergeometric Systems}},
  \href{https://doi.org/10.1007/JHEP04(2020)121}{\emph{JHEP} {\bfseries 04}
  (2020) 121} [\href{https://arxiv.org/abs/1910.08651}{{\ttfamily
  1910.08651}}].

\bibitem{delaCruz:2019skx}
L.~de~la Cruz, \emph{{Feynman integrals as A-hypergeometric functions}},
  \href{https://doi.org/10.1007/JHEP12(2019)123}{\emph{JHEP} {\bfseries 12}
  (2019) 123} [\href{https://arxiv.org/abs/1907.00507}{{\ttfamily
  1907.00507}}].

\bibitem{Schultka:2018nrs}
K.~Schultka, \emph{{Toric geometry and regularization of Feynman integrals}},
  \href{https://arxiv.org/abs/1806.01086}{{\ttfamily 1806.01086}}.

\bibitem{Schultka:2019tfi}
K.~Schultka, \emph{{Microlocal analyticity of Feynman integrals}}, Ph.D.
  thesis, Humboldt U., Berlin, 2019.
\newblock 10.18452/20161.

\bibitem{Primo:2016ebd}
A.~Primo and L.~Tancredi, \emph{{On the maximal cut of Feynman integrals and
  the solution of their differential equations}},
  \href{https://doi.org/10.1016/j.nuclphysb.2016.12.021}{\emph{Nucl. Phys.}
  {\bfseries B916} (2017) 94}
  [\href{https://arxiv.org/abs/1610.08397}{{\ttfamily 1610.08397}}].

\bibitem{Primo:2017ipr}
A.~Primo and L.~Tancredi, \emph{{Maximal cuts and differential equations for
  Feynman integrals. An application to the three-loop massive banana graph}},
  \href{https://doi.org/10.1016/j.nuclphysb.2017.05.018}{\emph{Nucl. Phys.}
  {\bfseries B921} (2017) 316}
  [\href{https://arxiv.org/abs/1704.05465}{{\ttfamily 1704.05465}}].

\bibitem{Bogner:2010kv}
C.~Bogner and S.~Weinzierl, \emph{{Feynman graph polynomials}},
  \href{https://doi.org/10.1142/S0217751X10049438}{\emph{Int. J. Mod. Phys. A}
  {\bfseries 25} (2010) 2585}
  [\href{https://arxiv.org/abs/1002.3458}{{\ttfamily 1002.3458}}].

\bibitem{Tarasov:1996br}
O.~Tarasov, \emph{{Connection between Feynman integrals having different values
  of the space-time dimension}},
  \href{https://doi.org/10.1103/PhysRevD.54.6479}{\emph{Phys. Rev. D}
  {\bfseries 54} (1996) 6479}
  [\href{https://arxiv.org/abs/hep-th/9606018}{{\ttfamily hep-th/9606018}}].

\bibitem{Lee:2009dh}
R.~Lee, \emph{{Space-time dimensionality D as complex variable: Calculating
  loop integrals using dimensional recurrence relation and analytical
  properties with respect to D}},
  \href{https://doi.org/10.1016/j.nuclphysb.2009.12.025}{\emph{Nucl. Phys. B}
  {\bfseries 830} (2010) 474}
  [\href{https://arxiv.org/abs/0911.0252}{{\ttfamily 0911.0252}}].

\bibitem{MR1269718}
V.~V. Batyrev, \emph{Dual polyhedra and mirror symmetry for {C}alabi-{Y}au
  hypersurfaces in toric varieties}, {\emph{J. Algebraic Geom.} {\bfseries 3}
  (1994) 493}.

\bibitem{Fanosearch}
T.~Coates, A.~Corti, S.~Galkin, V.~Golyshev and A.~Kasprzyk, ``Fano varieties
  and extremal laurent polynomials: A collaborative research blog.''
  {\tt{http://coates.ma.ic.ac.uk/fanosearch/}}, since December 2012.

\bibitem{Batyrev:1994pg}
V.~V. Batyrev and L.~A. Borisov, \emph{{On Calabi-Yau complete intersections in
  toric varieties}},  \href{https://arxiv.org/abs/alg-geom/9412017}{{\ttfamily
  alg-geom/9412017}}.

\bibitem{bailey2008elliptic}
D.~H. Bailey, J.~M. Borwein, D.~Broadhurst and M.~Glasser, \emph{Elliptic
  integral evaluations of bessel moments and applications}, {\emph{Journal of
  Physics A: Mathematical and Theoretical} {\bfseries 41} (2008) 205203}.

\bibitem{Broadhurst2013}
D.~Broadhurst, \emph{Multiple zeta values and modular forms in quantum field
  theory},  in \emph{Computer Algebra in Quantum Field Theory: Integration,
  Summation and Special Functions}, C.~Schneider and J.~Bl{\"u}mlein, eds.,
  (Vienna), pp.~33--73, Springer Vienna, (2013),
  \href{https://doi.org/10.1007/978-3-7091-1616-6_2}{DOI}.

\bibitem{Broadhurst:2018tey}
D.~Broadhurst and D.~P. Roberts, \emph{{Quadratic relations between Feynman
  integrals}}, \href{https://doi.org/10.22323/1.303.0053}{\emph{PoS} {\bfseries
  LL2018} (2018) 053}.

\bibitem{Zhou:2017jnm}
Y.~Zhou, \emph{{Wro\'nskian factorizations and Broadhurst--Mellit determinant
  formulae}}, \href{https://doi.org/10.4310/CNTP.2018.v12.n2.a5}{\emph{Commun.
  Num. Theor. Phys.} {\bfseries 12} (2018) 355}
  [\href{https://arxiv.org/abs/1711.01829}{{\ttfamily 1711.01829}}].

\bibitem{Zhou:2018tva}
Y.~Zhou, \emph{{Some algebraic and arithmetic properties of Feynman diagrams}},
   in \emph{{KMPB Conference}: {Elliptic Integrals, Elliptic Functions and
  Modular Forms in Quantum Field Theory}}, pp.~485--509, 2019,
  \href{https://doi.org/10.1007/978-3-030-04480-0\_19}{DOI}
  [\href{https://arxiv.org/abs/1801.05555}{{\ttfamily 1801.05555}}].

\bibitem{Zhou:2019rgc}
Y.~Zhou, \emph{{$\mathbb Q$-linear dependence of certain Bessel moments}},
  \href{https://arxiv.org/abs/1911.04141}{{\ttfamily 1911.04141}}.

\bibitem{Broadhurst:2016myo}
D.~Broadhurst, \emph{{Feynman integrals, L-series and Kloosterman moments}},
  \href{https://doi.org/10.4310/CNTP.2016.v10.n3.a3}{\emph{Commun. Num. Theor.
  Phys.} {\bfseries 10} (2016) 527}
  [\href{https://arxiv.org/abs/1604.03057}{{\ttfamily 1604.03057}}].

\bibitem{fresn2018hodge}
J.~Fres{\'a}n, C.~Sabbah and J.-D. Yu, \emph{{Hodge theory of Kloosterman
  connections}},  \href{https://arxiv.org/abs/1810.06454}{{\ttfamily
  1810.06454}}.

\bibitem{fresn2020quadratic}
J.~Fres{\'a}n, C.~Sabbah and J.-D. Yu, \emph{{Quadratic relations between
  Bessel moments}},  \href{https://arxiv.org/abs/2006.02702}{{\ttfamily
  2006.02702}}.

\bibitem{verrill2004sums}
H.~A. Verrill, \emph{{Sums of squares of binomial coefficients, with
  applications to Picard-Fuchs equations}},
  \href{https://arxiv.org/abs/math/0407327}{{\ttfamily math/0407327}}.

\bibitem{Borwein_2008}
J.~M. Borwein and B.~Salvy, \emph{A proof of a recurrence for bessel moments},
  \href{https://doi.org/10.1080/10586458.2008.10129032}{\emph{Experimental
  Mathematics} {\bfseries 17} (2008) 223}.

\bibitem{MR1809982}
M.~Bronstein, T.~Mulders and J.-A. Weil, \emph{On symmetric powers of
  differential operators},  in \emph{Proceedings of the 1997 {I}nternational
  {S}ymposium on {S}ymbolic and {A}lgebraic {C}omputation ({K}ihei, {HI})},
  pp.~156--163, ACM, New York, 1997,
  \href{https://doi.org/10.1145/258726.258771}{DOI}.

\bibitem{MR1689204}
A.~Libgober, \emph{Chern classes and the periods of mirrors},
  \href{https://doi.org/10.4310/MRL.1999.v6.n2.a2}{\emph{Math. Res. Lett.}
  {\bfseries 6} (1999) 141}.

\bibitem{MR1838444}
S.~Hosono, \emph{Local mirror symmetry and type {IIA} monodromy of
  {C}alabi-{Y}au manifolds},
  \href{https://doi.org/10.4310/ATMP.2000.v4.n2.a5}{\emph{Adv. Theor. Math.
  Phys.} {\bfseries 4} (2000) 335}.

\bibitem{MR2683208}
H.~Iritani, \emph{Ruan's conjecture and integral structures in quantum
  cohomology},  in \emph{New developments in algebraic geometry, integrable
  systems and mirror symmetry ({RIMS}, {K}yoto, 2008)}, vol.~59 of \emph{Adv.
  Stud. Pure Math.}, pp.~111--166, Math. Soc. Japan, Tokyo, (2010),
  \href{https://doi.org/10.2969/aspm/05910111}{DOI}.

\bibitem{MR2483750}
L.~Katzarkov, M.~Kontsevich and T.~Pantev, \emph{Hodge theoretic aspects of
  mirror symmetry},  in \emph{From {H}odge theory to integrability and {TQFT}
  tt*-geometry}, vol.~78 of \emph{Proc. Sympos. Pure Math.}, pp.~87--174, Amer.
  Math. Soc., Providence, RI, (2008),
  \href{https://doi.org/10.1090/pspum/078/2483750}{DOI}.

\bibitem{MR3536989}
S.~Galkin, V.~Golyshev and H.~Iritani, \emph{Gamma classes and quantum
  cohomology of {F}ano manifolds: gamma conjectures},
  \href{https://doi.org/10.1215/00127094-3476593}{\emph{Duke Math. J.}
  {\bfseries 165} (2016) 2005}.

\bibitem{Klemm:1996ts}
A.~Klemm, B.~Lian, S.~Roan and S.-T. Yau, \emph{{Calabi-Yau fourfolds for M
  theory and F theory compactifications}},
  \href{https://doi.org/10.1016/S0550-3213(97)00798-0}{\emph{Nucl. Phys. B}
  {\bfseries 518} (1998) 515}
  [\href{https://arxiv.org/abs/hep-th/9701023}{{\ttfamily hep-th/9701023}}].

\bibitem{Bizet:2014uua}
N.~Cabo~Bizet, A.~Klemm and D.~Vieira~Lopes, \emph{{Landscaping with fluxes and
  the E8 Yukawa Point in F-theory}},
  \href{https://arxiv.org/abs/1404.7645}{{\ttfamily 1404.7645}}.

\bibitem{MR1228584}
K.~Oguiso, \emph{On algebraic fiber space structures on a {C}alabi-{Y}au
  {$3$}-fold}, \href{https://doi.org/10.1142/S0129167X93000248}{\emph{Internat.
  J. Math.} {\bfseries 4} (1993) 439}.

\bibitem{privcomiritani}
H.~Iritani, \emph{Private communication}, .

\bibitem{Matt:progress}
M.~Kerr, \emph{Work in progress}, .

\bibitem{Zhou:2017vhw}
Y.~Zhou, \emph{{Wick rotations, Eichler integrals, and multi-loop Feynman
  diagrams}}, \href{https://doi.org/10.4310/CNTP.2018.v12.n1.a5}{\emph{Commun.
  Num. Theor. Phys.} {\bfseries 12} (2018) 127}
  [\href{https://arxiv.org/abs/1706.08308}{{\ttfamily 1706.08308}}].

\bibitem{Bloch:2014qca}
S.~Bloch, M.~Kerr and P.~Vanhove, \emph{{A Feynman integral via higher normal
  functions}}, \href{https://doi.org/10.1112/S0010437X15007472}{\emph{Compos.
  Math.} {\bfseries 151} (2015) 2329}
  [\href{https://arxiv.org/abs/1406.2664}{{\ttfamily 1406.2664}}].

\bibitem{MR1184764}
C.~Peters, J.~Top and M.~van~der Vlugt, \emph{The {H}asse zeta function of a
  {$K3$} surface related to the number of words of weight {$5$} in the {M}elas
  codes}, {\emph{J. Reine Angew. Math.} {\bfseries 432} (1992) 151}.

\bibitem{hulek2005modularity}
K.~Hulek and H.~Verrill, \emph{On modularity of rigid and nonrigid calabi-yau
  varieties associated to the root lattice a 4}, {\emph{Nagoya Mathematical
  Journal} {\bfseries 179} (2005) 103}.

\bibitem{Candelas:2019llw}
P.~Candelas, X.~de~la Ossa, M.~Elmi and D.~Van~Straten, \emph{{A One Parameter
  Family of Calabi-Yau Manifolds with Attractor Points of Rank Two}},
  \href{https://arxiv.org/abs/1912.06146}{{\ttfamily 1912.06146}}.

\bibitem{BoenischThesis}
K.~B{\"o}nisch, \emph{Modularity, periods and quasiperiods at special points in
  calabi-yau moduli spaces},  Master's thesis, University of Bonn, 2020,
  \url{http://www.th.physik.uni-bonn.de/Groups/Klemm/data.php}.

\bibitem{Laporta:2004rb}
S.~Laporta and E.~Remiddi, \emph{{Analytic treatment of the two loop equal mass
  sunrise graph}},
  \href{https://doi.org/10.1016/j.nuclphysb.2004.10.044}{\emph{Nucl. Phys. B}
  {\bfseries 704} (2005) 349}
  [\href{https://arxiv.org/abs/hep-ph/0406160}{{\ttfamily hep-ph/0406160}}].

\bibitem{Caffo:1998du}
M.~Caffo, H.~Czyz, S.~Laporta and E.~Remiddi, \emph{{The Master differential
  equations for the two loop sunrise selfmass amplitudes}}, {\emph{Nuovo Cim.
  A} {\bfseries 111} (1998) 365}
  [\href{https://arxiv.org/abs/hep-th/9805118}{{\ttfamily hep-th/9805118}}].

\bibitem{Schnetz:2016fhy}
O.~Schnetz, \emph{{Numbers and Functions in Quantum Field Theory}},
  \href{https://doi.org/10.1103/PhysRevD.97.085018}{\emph{Phys. Rev. D}
  {\bfseries 97} (2018) 085018}
  [\href{https://arxiv.org/abs/1606.08598}{{\ttfamily 1606.08598}}].

\bibitem{Itzykson:1980rh}
C.~Itzykson and J.~Zuber, \emph{{Quantum Field Theory}}, International Series
  In Pure and Applied Physics. McGraw-Hill, New York, 1980.

\bibitem{Brown:2012ia}
F.~Brown and O.~Schnetz, \emph{{Proof of the zig-zag conjecture}},
  \href{https://arxiv.org/abs/1208.1890}{{\ttfamily 1208.1890}}.

\bibitem{Brown:2013wn}
F.~Brown and D.~Doryn, \emph{{Framings for graph hypersurfaces}},
  \href{https://arxiv.org/abs/1301.3056}{{\ttfamily 1301.3056}}.

\bibitem{Brown:2010bw}
F.~Brown and O.~Schnetz, \emph{{A K3 in $\phi^4$}},
  \href{https://doi.org/10.1215/00127094-1644201}{\emph{Duke Math. J.}
  {\bfseries 161} (2012) 1817}
  [\href{https://arxiv.org/abs/1006.4064}{{\ttfamily 1006.4064}}].

\bibitem{Panzer:2016snt}
E.~Panzer and O.~Schnetz, \emph{{The Galois coaction on $\phi^4$ periods}},
  \href{https://doi.org/10.4310/CNTP.2017.v11.n3.a3}{\emph{Commun. Num. Theor.
  Phys.} {\bfseries 11} (2 017) 657}
  [\href{https://arxiv.org/abs/1603.04289}{{\ttfamily 1603.04289}}].

\bibitem{Schnetz:2017bko}
O.~Schnetz, \emph{{The Galois coaction on the electron anomalous magnetic
  moment}}, \href{https://doi.org/10.4310/CNTP.2018.v12.n2.a4}{\emph{Commun.
  Num. Theor. Phys.} {\bfseries 12} (2018) 335}
  [\href{https://arxiv.org/abs/1711.05118}{{\ttfamily 1711.05118}}].

\bibitem{Laporta:2017okg}
S.~Laporta, \emph{{High-precision calculation of the 4-loop contribution to the
  electron g-2 in QED}},
  \href{https://doi.org/10.1016/j.physletb.2017.06.056}{\emph{Phys. Lett. B}
  {\bfseries 772} (2017) 232}
  [\href{https://arxiv.org/abs/1704.06996}{{\ttfamily 1704.06996}}].

\bibitem{Huang:2014nwa}
M.-x. Huang, A.~Klemm, J.~Reuter and M.~Schiereck, \emph{{Quantum geometry of
  del Pezzo surfaces in the Nekrasov-Shatashvili limit}},
  \href{https://doi.org/10.1007/JHEP02(2015)031}{\emph{JHEP} {\bfseries 02}
  (2015) 031} [\href{https://arxiv.org/abs/1401.4723}{{\ttfamily 1401.4723}}].

\bibitem{Fischbach:2018yiu}
F.~Fischbach, A.~Klemm and C.~Nega, \emph{{WKB Method and Quantum Periods
  beyond Genus One}}, \href{https://doi.org/10.1088/1751-8121/aae8b0}{\emph{J.
  Phys. A} {\bfseries 52} (2019) 075402}
  [\href{https://arxiv.org/abs/1803.11222}{{\ttfamily 1803.11222}}].

\bibitem{Bloch:2013tra}
S.~Bloch and P.~Vanhove, \emph{{The elliptic dilogarithm for the sunset
  graph}}, \href{https://doi.org/10.1016/j.jnt.2014.09.032}{\emph{J. Number
  Theor.} {\bfseries 148} (2015) 328}
  [\href{https://arxiv.org/abs/1309.5865}{{\ttfamily 1309.5865}}].

\bibitem{Klemm:1995tj}
A.~Klemm, W.~Lerche and P.~Mayr, \emph{{K3 Fibrations and heterotic type II
  string duality}},
  \href{https://doi.org/10.1016/0370-2693(95)00937-G}{\emph{Phys. Lett. B}
  {\bfseries 357} (1995) 313}
  [\href{https://arxiv.org/abs/hep-th/9506112}{{\ttfamily hep-th/9506112}}].

\bibitem{Marino:1998pg}
M.~Marino and G.~W. Moore, \emph{{Counting higher genus curves in a Calabi-Yau
  manifold}}, \href{https://doi.org/10.1016/S0550-3213(98)00847-5}{\emph{Nucl.
  Phys. B} {\bfseries 543} (1999) 592}
  [\href{https://arxiv.org/abs/hep-th/9808131}{{\ttfamily hep-th/9808131}}].

\bibitem{Klemm:2005pd}
A.~Klemm and M.~Marino, \emph{{Counting BPS states on the enriques
  Calabi-Yau}}, \href{https://doi.org/10.1007/s00220-007-0407-z}{\emph{Commun.
  Math. Phys.} {\bfseries 280} (2008) 27}
  [\href{https://arxiv.org/abs/hep-th/0512227}{{\ttfamily hep-th/0512227}}].

\bibitem{Grimm:2007tm}
T.~W. Grimm, A.~Klemm, M.~Marino and M.~Weiss, \emph{{Direct Integration of the
  Topological String}},
  \href{https://doi.org/10.1088/1126-6708/2007/08/058}{\emph{JHEP} {\bfseries
  08} (2007) 058} [\href{https://arxiv.org/abs/hep-th/0702187}{{\ttfamily
  hep-th/0702187}}].

\end{thebibliography}\endgroup







\end{document}